\documentclass[conference]{IEEEtran}
\usepackage[pdftex]{graphicx}
\usepackage{listings}
\usepackage{graphicx}
\usepackage{caption}
\usepackage{subcaption}
\usepackage{amsmath}
\usepackage{setspace}
\usepackage{algorithm}
\usepackage{algorithmic}
\usepackage{epstopdf}
\usepackage[pdftex]{graphicx}
\usepackage{amssymb}
\usepackage{authblk}
\usepackage{diagbox}
\usepackage{threeparttable}
\usepackage{graphicx}
\usepackage{booktabs}
\usepackage{chngcntr}
\usepackage{multirow}
\usepackage{cite}
\usepackage{float}
\usepackage{threeparttable}
\usepackage{amssymb}
\newcommand{\norm}[1]{\left\lVert#1\right\rVert}

\ifCLASSINFOpdf
\else
\fi
\begin{document}

\title{Unsupervised Learning of Spike Patterns for Seizure Detection and Wavefront Estimation of High Resolution Micro Electrocorticographic ($\mu$ECoG) Data }
%Seizure Detection Through Manifold Clustering and Temporal Analysis of Micro Electrocorticographic ($\mu$ECoG) Data } 

%\author{\IEEEauthorblockN{Yilin Song}
%\IEEEauthorblockA{School of Electrical and\\Computer Engineering\\
%Polytechnic Institute of New York University\\
%Email: ys1297@nyu.edu}
%\and
%\IEEEauthorblockN{Bugra Akyildiz}
%\IEEEauthorblockA{School of Electrical and\\Computer Engineering\\
%Polytechnic Institute of New York University\\
%Email: ba830@nyu.edu}
%\and
%\IEEEauthorblockN{Jonathan Viventi}
%\IEEEauthorblockA{School of Electrical and\\Computer Engineering\\
%Polytechnic Institute of New York University\\
%Emai:jviventi@nyu.edu}
%\and
%\IEEEauthorblockN{Yao Wang}
%\IEEEauthorblockA{School of Electrical and\\Computer Engineering\\
%Polytechnic Institute of New York University\\
%Email:yao@nyu.edu}}

\author[1]{Yilin Song }
\author[2]{Jonathan Viventi }
\author[1]{Yao Wang }

\affil[1]{Department of Electrical and Computer Engineering,  New York University, NY, USA}
\affil[2]{Department of Biomedical Engineering, Duke University, Durham, NC, USA}


%Telephone: (800) 555--1212, Fax: (888) 555--1212}


\maketitle

\begin{abstract}
For the past few years, we have developed flexible, active, multiplexed recording devices for high resolution recording over large, clinically relevant areas in the brain. While this technology has enabled a much higher-resolution view of the electrical activity of the brain, the analytical methods to process, categorize and respond to the huge volumes of seizure data produced by these devices have not yet been developed. In this work we proposed an unsupervised learning framework for spike analysis, which by itself reveals spike pattern. By applying advanced video processing techniques for separating a multi-channel recording into individual spike segments, unfolding the spike segments manifold and identifying natural clusters for spike patterns, we are able to find the common spike motion patterns. And we further explored using these patterns for more interesting and practical problems as seizure prediction and spike wavefront prediction. These methods have been applied to in-vivo feline seizure recordings and yielded promising results. 

\end{abstract}

\IEEEpeerreviewmaketitle

\section{introduction}
Currently many existing neurological data analyses rely on manual inspection. With new high-density electrode arrays that provide dramatically enhanced spatial resolution, the data volume is too large for manual review. Further, manual inspection can miss subtle features that automated machine learning techniques can detect. There is an urgent need for efficient and sensitive automated methods that can analyze the large volumes of data produced by next generation neurologic devices. 

One major application of high resolution Micro Electrocorticographic ($\mu$ECoG) is to record brain signals in patients with epilepsy. In our work, a spike segment refers to a consecutive set of frames in which one or several adjacent channels have high amplitude. We believe different spike patterns correspond to different neuron states, a thorough study of the wave patterns of spike would give better understanding towards the true nature of neuron activities. Given the fact that there is almost no way for human to label the data in precise time point or even a short time horizon, a fully unsupervised technique to analyze the data is crucial. We propose an unsupervised spike clustering framework to reveal activities within and between spike segments. And further experiments validate that this clustering method helped to achieve wave front prediction and seizure detection.

 In previous work \cite{delay_map1} we used human inspection to find bad channels and gaussian spatial filter to smooth each frame. These methods could bring in blocky artifacts in filtered frames. And these artifacts would further affect the accuracy of spike segmentation. To overcome this, we first developed a graph based signal processing framework for our dataset. Graph filtering serves the purpose of automatically locate bad channels of our $\mu$ECoG data either caused by manufacturing defects or loss of contact between channel and tissue. It also removes noisy spatial patterns and improves performance of spike segmentation.  

For spike segmentation, we treat the multi-channel signal as a 3-D volume (successive 2D image frames in time) and apply a 3D region growing technique to determine the spike region. Region growing \cite{Region_Growing} is an effective technique to detect a connected (in 3D) region in which each pixel has an intensity that is noticeably higher (or lower) than pixels outside the region. The reason we prefer region growing based techniques to fixed length threshold crossing segmentation as we used in \cite{delay_map1} is that each spike may last for somewhat different duration depending on its initial location and its moving speed. Fixed length segmentation scheme sometimes includes more than one spike or an incomplete spike in a detected segment.

\begin{figure}[h]
        \centering
                %\centering
                \includegraphics[width=0.5\textwidth]{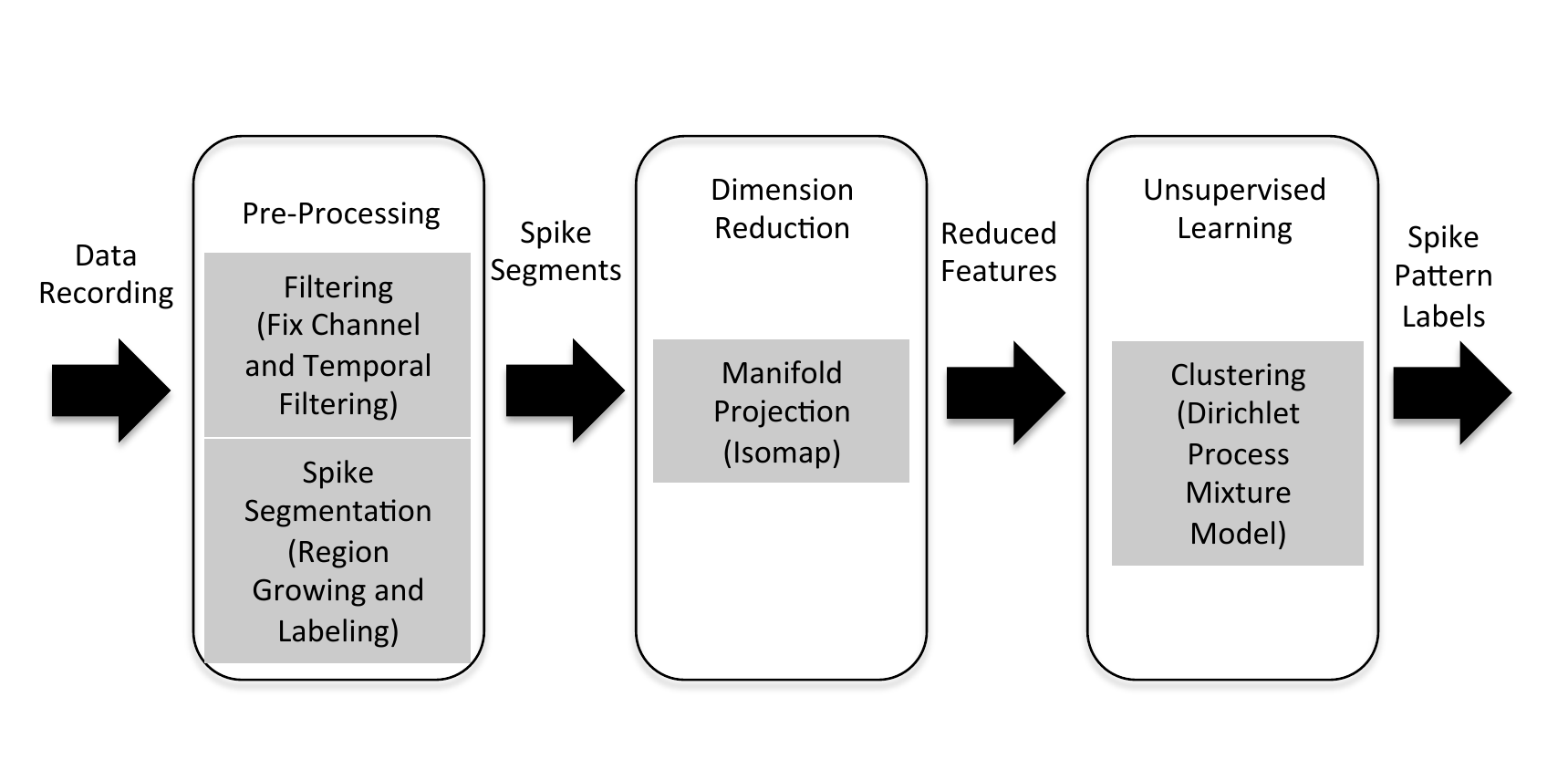}
                \caption{ \textit{\small{Processing steps for clustering spike patterns.}}}
                \label{Block_diagram_1}
\end{figure}

The detected electrographic spikes live in a high dimensional space defined by electrode array layout and temporal resolution. Any clustering or classification algorithm directly applied to these spike segments would suffer from the curse of dimensionality. We build a pairwise metric for spike segments to model the manifold spike segments live in. Recently we have compared several clustering algorithms on different features in \cite{akyildiz2013improved}. In here we mainly focus on how we construct the manifold and how we unveil the structure using Isomap\cite{Isomap}. We exploit Dirichlet Process Mixture Model (DPM) for clustering. The benefits of DPM is that it can automatically determine the number of clusters. The overall spike pattern clustering involves three layers as shown in Fig. \ref{Block_diagram_1}. 

One applications of the spike clustering is wavefront prediction. The wavefront of a spike is ideally the leading edge of the moving region. For example, for a square region moving to the right, the right edge of the square would be considered the wavefront. For our discussion, the wavefront at each frame refers to an intermediate point inside spike region that has a strong signal and is moving in the same direction as the whole region (more precise definition is given in Section. \ref{Delay Map and Wavefront Trajectory}. Because the wavefront trajectory varies from one spike to another in a highly non-linear way, wavefront prediction is an extremely challenging task. To the best of our knowledge, we are the first to attempt to solve this problem. 

In addition to wavefront prediction, we also investigate if the spike patterns could be used for seizure detection and prediction. Most current seizure detection/prediction methods rely on clever designed features like spectral power, wavelet energy, spike rate and so on\cite{bandarabadi2015epileptic,li2013seizure,eftekhar2014ngram,gadhoumi2012discriminating}. Whereas in our case, we predict seizure based on the spike label variation in time which is modeled by hidden markov model (HMM). One straightforward approach is using the HMM hidden state to directly model the seizure stage, but we found that there are limited seizures in our dataset to learn the correct transition probability between hidden states. Hence we built three HMMs corresponding to three seizure stages (non-seizure, pre-seizure, and seizure), and let each individual HMM captures the temporal dynamics of that particular seizure stage.
%
%We classified spikes into one of the three seizure stages (non-seizure, pre-seizure, and seizure). A spike that is classified into the pre-seizure stage indicates the onset of a new seizure. We first labeled each spike into one of the pre-determined spike clusters based on features extracted from this spike signal. Then we classified each spike based on how the spike pattern changes in $T$ successive spikes surrounding the current one. This is accomplished through HMM modeling of the dynamics of spike labels in different seizure stages. 

The rest of this paper is organized as follows:  In Section \ref{signal processing as a graph}, we present the algorithm for preprocessing using graph filtering. In Section \ref{spike segmentation}, we illustrate spike segmentation using region growing techniques. In Section \ref{spike clustering through manifold learning}, we introduce the framework for spike pattern clustering. In Section \ref{TRAJECTORY PREDICTION} and Section \ref{Seizure Detection and Prediction} we demonstrate how this clustering technique helps to solve problems including wavefront prediction and seizure detection. Section \ref{conclusion} concludes the paper with a discussion and future work.

\section{preprocessing using graph filtering} \label{signal processing as a graph}
 \begin{figure}
        \centering
        \begin{subfigure}[b]{0.225\textwidth}
                %\centering
                \includegraphics[width=\textwidth]{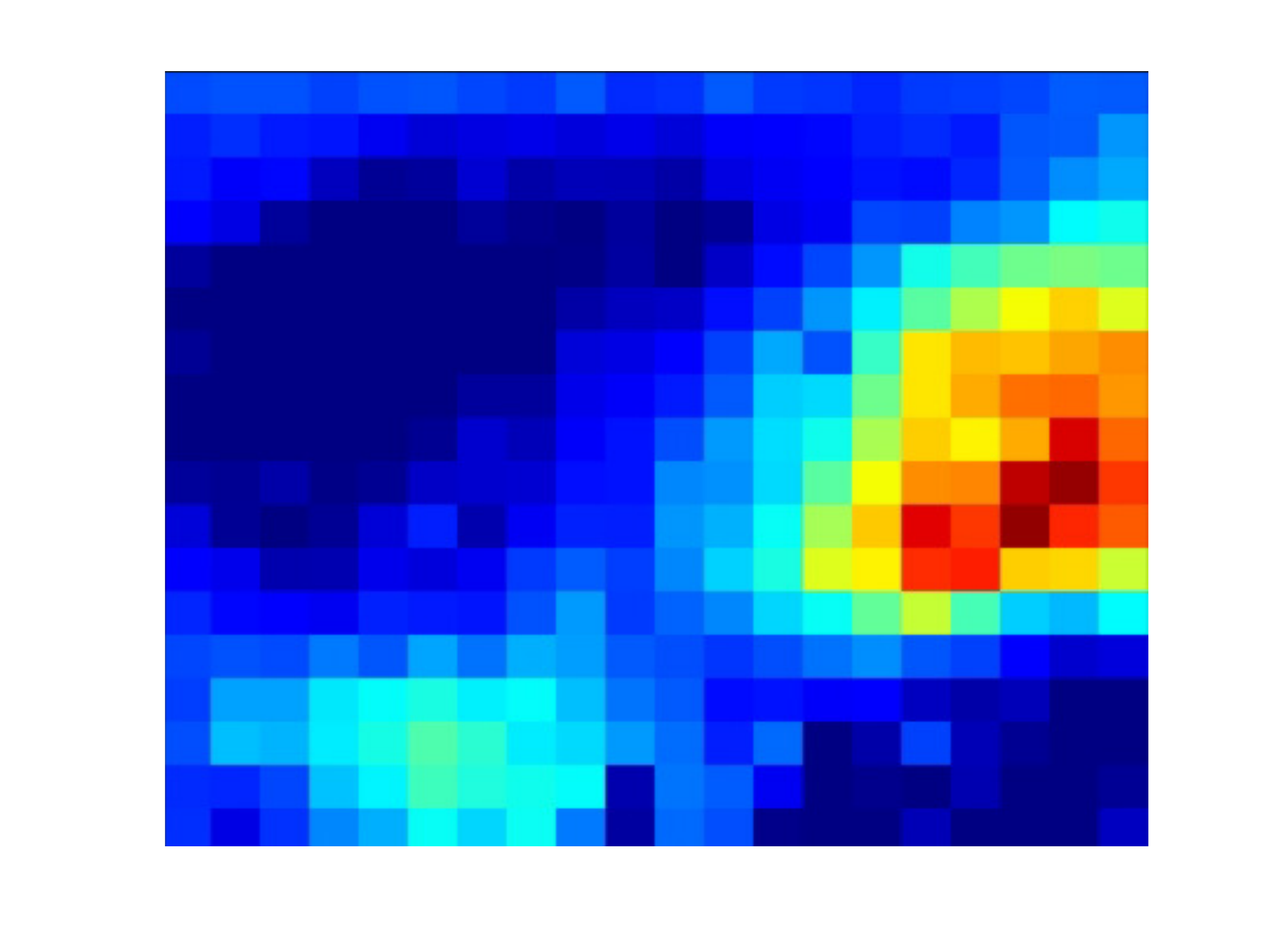}
                \caption{ }
          \end{subfigure}%
	\begin{subfigure}[b]{0.225\textwidth}
                %\centering
                \includegraphics[width=\textwidth]{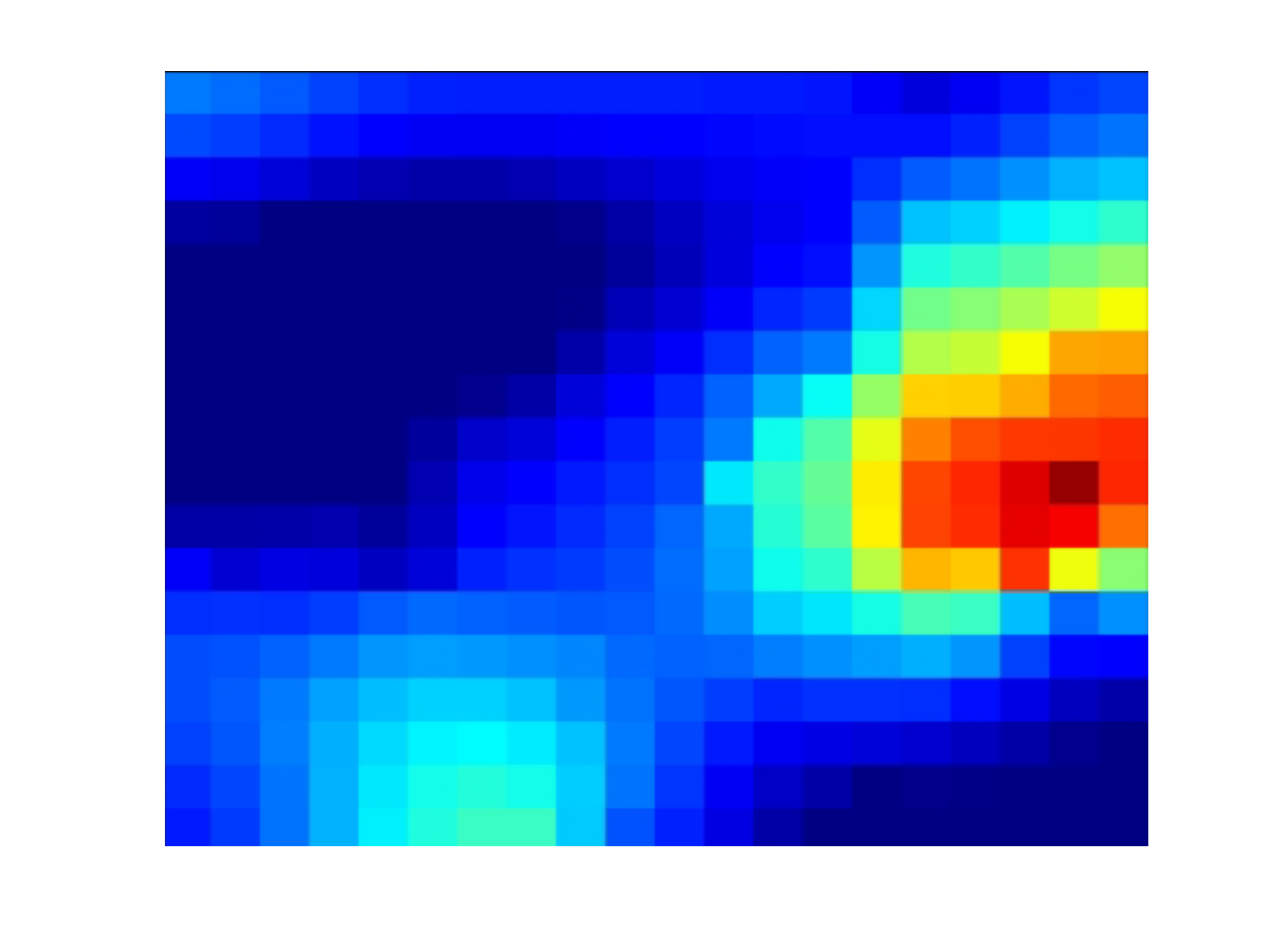}
                \caption{}
                      \end{subfigure}%
        
        ~
        
	%add desired spacing between images, e. g. ~, \quad, \qquad etc.
        \begin{subfigure}[b]{0.225\textwidth}
                %\centering
                \includegraphics[width=\textwidth]{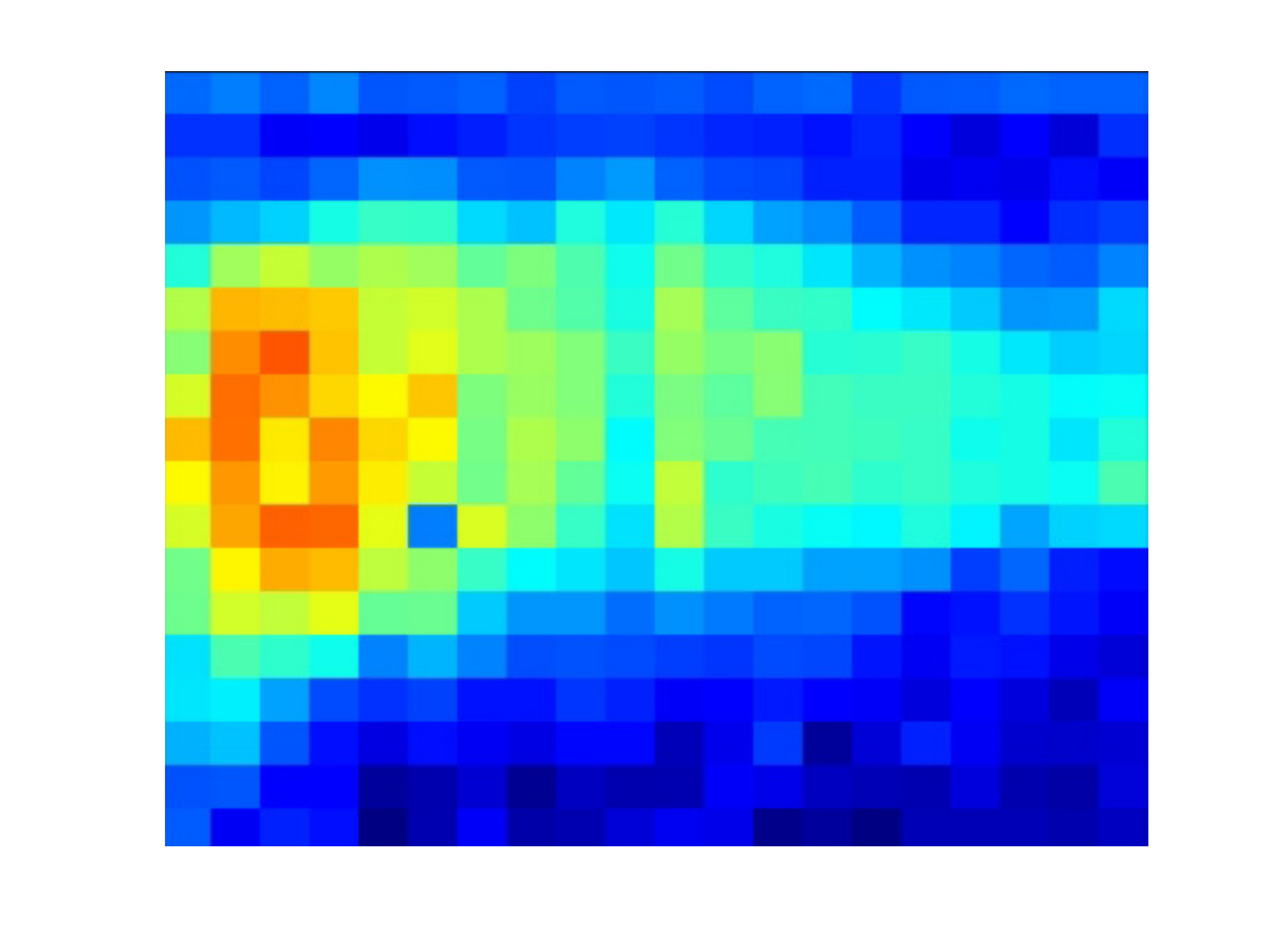}
                \caption{}
                       \end{subfigure}
	\begin{subfigure}[b]{0.225\textwidth}
                %\centering
                \includegraphics[width=\textwidth]{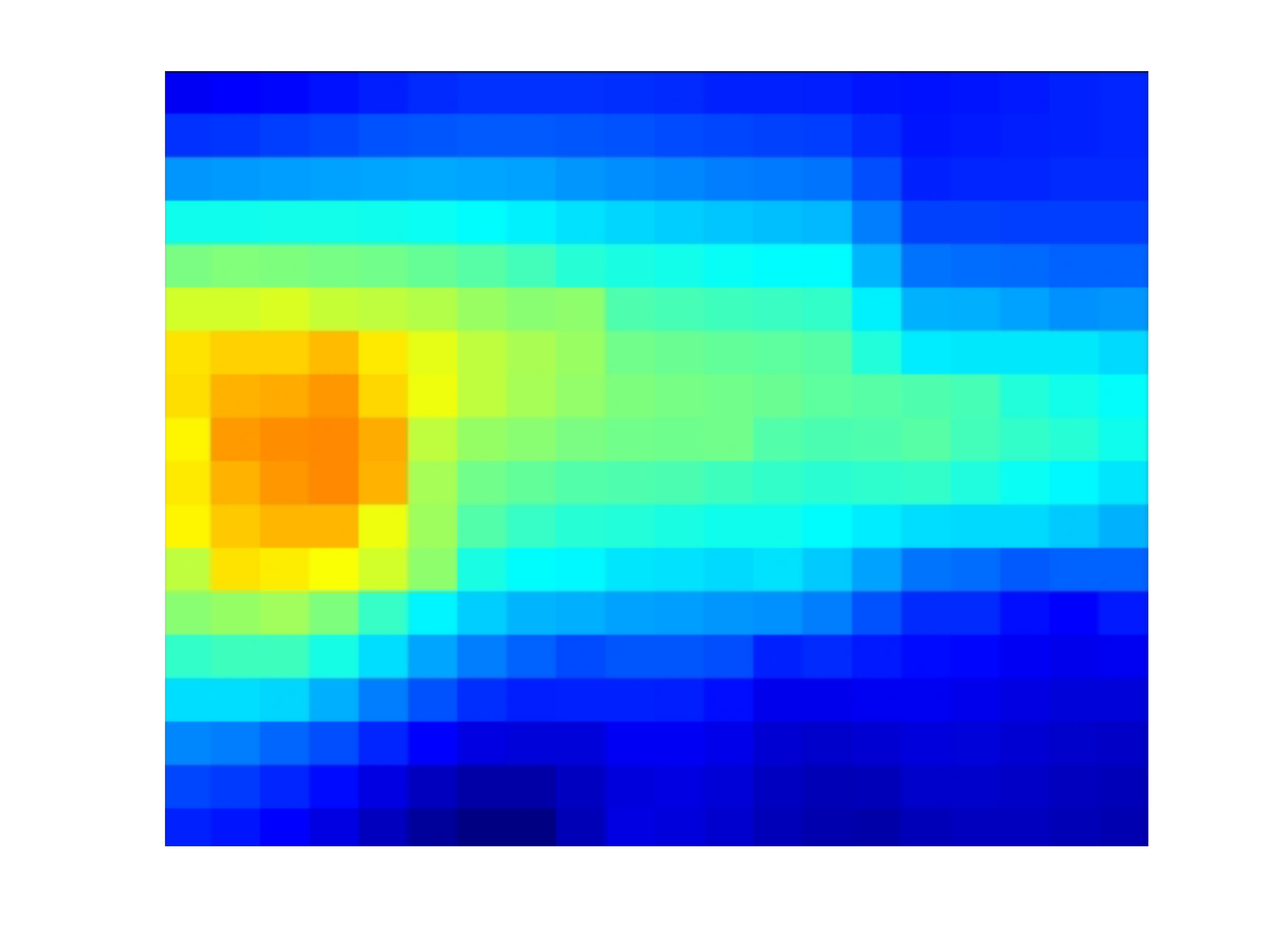}
                \caption{}
            
        \end{subfigure}
        \caption{\textit{\small{Comparison of graph filtering and Gaussian smoothing. Subplots a,c are results using a 5 by 5 Gaussian smoothing kernel for spatial filtering, and subplots b,d are graph filtering results using 80 dominant graph coefficients. For Gaussian filtering bad channels are human labelled and interpolated with Gaussian distance kernel from good channels within a local region of 5 by 5. The results b,d preserved the original pattern as well as smoothed the signal. }}}
                
        %\caption{Comparison of fixed length segmentation scheme against using variable length region based segmentation method. Subfigure b,d is using variable length region based segmentation method and Subfigure a,c is using fixed length segmentation method as described in \cite{delay_map1}. Subfigure a is showing a spike is falsely segmented by fixed length segmentation method, where as Subfigure c is showing spikes being falsely connected. The 3D volume in subfigure a,c are the high intensity pixels, where in fact the entire volume is considered as spike with fix length segmentation. In subfigure d, different spike segments are represented with different colors}
        
 \label{graph filtering}
\end{figure}

All channel recordings are individually band-pass filtered between 1 and 50 Hz with a 6th order butterworth filter in the forward and reverse direction. This is to remove the high frequency noise in the temporal domain. In order to remove the noise in the spatial domain and fill in missing channels caused by manufacturing defects or lose of contact on human membrane, we developed graph filtering technique. Specifically, we represent a $\mu$ECOG dataset as a graph, where each vertex of the graph represents one signal channel, and the edge between two vertices embeds the relational information between two channels. The weighted graph $G =\{V,E,W\}$ consists  of a finite set of vertices and a weighted adjacency matrix $W$. We want the adjacency matrix $W$ not only reflects the spatial relationship between two channels but also reflects the signal correlations. The adjacency matrix is defined as:

\begin{equation}
\begin{aligned}
W(i,j)&= D(i,j) *C(i,j)\\
D(i,j)& = \Bigg\{ 
\begin{aligned}
&exp(- ( \frac{dist(i,j)^2}{2\theta^2})) &  if \quad dist(i,j)^2 \leq 32\\
&0 &otherwise\\
\end{aligned}\\
C(i,j)& = \Bigg\{
\begin{aligned}
&Corr(i,j)\quad & if   \quad Corr(i,j) \geq 0.8 \\
&0 &otherwise\\
\end{aligned}
\label{graph_adjacency}
\end{aligned}
%\footnotesize{
%W_{i,j} = \Bigg\{ 
%\begin{aligned}
%& Dist(i,j)* Corr(i,j) & if \quad dist(i,j) \leq k \quad and \quad Corr(i,j) \geq 0.9\\
%& 0 & otherwise \\
%\end{aligned}}
\end{equation}
%  exp(- ( \frac{dist(i,j)^2}{2\theta^2}))

In here, the distance $dist(i,j)$ is the Euclidean distance in the $x,y$ coordinates between channel $i$ and $j$. 
The hard thresholding of distance and correlation aims to generate a sparse adjacency matrix and potentially a disconnected graph. We have experienced with various threshold values, and through trial and error, we found that setting distance and correlation threshold at 32 and 0.8, respectively, generates the disconnected graph quite well.
The disconnected graph serves the purpose of filtering out the noisy channels caused by manufacturing defects or lose of contact and relieves the burden of human inspection. If all channels work properly, then the graph $G=\{V,E,W\}$ would be a connected one. Otherwise, the graph would have a bunch of disconnected subgraphs with each subgraph consisting of the electrode channels with strong connections. The largest connected subgraph represents the good channels as we assume the majority of our electrode array are functioning properly and are correlated with each other. Then the set of bad channels are interpolated from good channels as $f_{bad} = M f_{out}$. Here $f_{bad}$ at any frame is a vector consisting of the signals of bad channels, $f_{out}$ is a vector consisting of the filtered signals of good channels, and M is a matrix with elements defined in Eq (2), in which $i$ stands for one bad channel, $j$ stands for a good channel.  
%Then each bad channel $i$ then is interpolated with a linear combination of the set of reconstructed good channels.
%The malfunctioning channels are then filled in based on the spatial relationship between the good channels using the graph coefficients we get from Eq. \ref{graph_coefficeint}.
\begin{equation}
M(i,j) = \frac{exp(- ( \frac{dist(i,j)^2}{2\theta^2}))} { \sum_j(exp(- ( \frac{dist(i,j)^2}{2\theta^2})))} 
\label{bad_channel}
\end{equation}

To filter the good channel signals, we form the non-normalized graph laplacian of the largest connected subgraph $G_{sub}$ as $L_{G_{sub}} = D_{G_{sub}}-W_{G_{sub}}$. $W_{G_{sub}}$ is the adjacency matrix for the subgraph, and the degree matrix $D_{G_{sub}}$ is a diagonal matrix whose diagonal element in each row is the sum of all the weights in the same row of matrix $W_{G_{sub}}$. Because graph Laplacian matrix $L_{G_{sub}}$ is a real symmetric matrix, it has a set of orthonormal eigenvectors represented by $U = \{u_n\}_{n =0,1,\cdots,N-1}$, where N is the number of good channels. Since the graph is connected, the ordered eigenvalue associated with the eigenvectors has the following traits $0 = \lambda_0 \le \lambda_1 \leq \lambda_2 \cdots \leq \lambda_{N-1}$. The spectrum of the graph laplacian is defined as $\sigma(L) \equiv \{\lambda_0, \lambda_1, \cdots, \lambda_{N-1}\}$.
Once we have a graph spectral representation, in order to filter out high frequency component in the spatial domain, we can generalize filtering in signal processing to graph spectral filtering \cite{shuman2013emerging} as:
\begin{equation}
\begin{aligned}
&f_{out} = \hat{h}(L_{G_{sub}}) f_{in} \\
& \hat{h}(L) = U \begin{pmatrix} 
\hat{h}(\lambda_0)& \quad & 0 \\
\quad & \ddots &\quad \\
0 & \quad & \hat{h}(\lambda_{N-1}) 
\end{pmatrix} U^T\\
\label{graph_coefficeint}
%\hat{f}_{out}(\lambda_l) = \hat{f}_{in}(\lambda_l) \hat{h}(\lambda_l) 
\end{aligned}
\end{equation}

%In Eq. \ref{graph_coefficeint}, the set of unfiltered good channel signals at one particular time point are represented as $f_{in}$. And the filtered good channels are represented as $f_{out}$. 
In our case, a simple thresholding of the graph spectrum could render good reconstruction result. This is done by setting $\hat{h}(\lambda_{t}) = \lambda_{t}$ for $t \leq T$  and 0 otherwise. It resembles low pass filtering in traditional signal processing. The threshold $T$ is selected to be the smallest number to preserve at least $90\%$ of original energy. Figure \ref{graph filtering} shows a comparison between graph filtering and gaussian spatial filtering. The blocking artifact caused by Gaussian filtering is removed by using graph filtering.

\section{spike segmentation} \label{spike segmentation} 

\begin{figure}
        \centering
        \begin{subfigure}[b]{0.15\textwidth}
                %\centering
                \includegraphics[width=\textwidth]{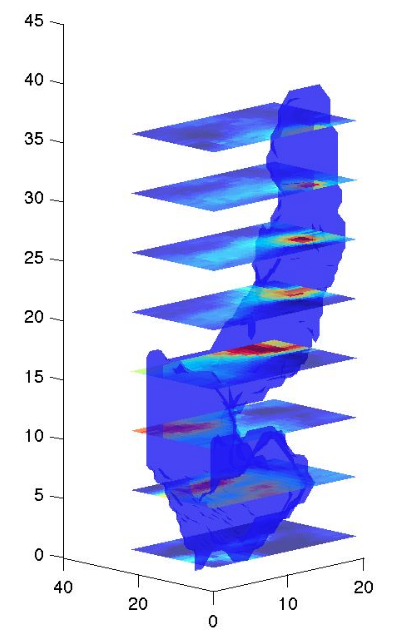}
                \caption{}
                \label{spiral}
        \end{subfigure}%
%	\begin{subfigure}[b]{0.15\textwidth}
%                %\centering
%                \includegraphics[width=\textwidth]{spiral_T.jpg}
%                \caption{}
%                \label{spiral traject}
%        \end{subfigure}%
%
%        ~ 
	%add desired spacing between images, e. g. ~, \quad, \qquad etc.
        \begin{subfigure}[b]{0.15\textwidth}
                %\centering
                \includegraphics[width=\textwidth]{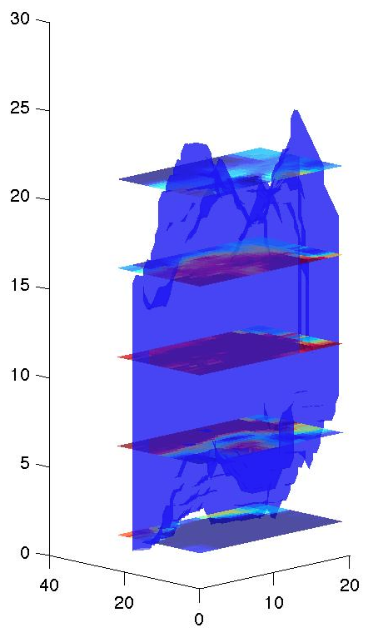}
                \caption{}
                \label{plane}
        \end{subfigure}
%	\begin{subfigure}[b]{0.15\textwidth}
%                %\centering
%                \includegraphics[width=\textwidth]{plane_T.jpg}
%                \caption{}
%                \label{plane traject}
%        \end{subfigure}%
        \caption{\textit{\small{Examples of spike segmentation. Subfigure a is for a spike with a spiral motion. Subfigure b is for a spike with a planar motion.
        Each subfigure shows the segmented volume in transparent blue overlaid with the $\mu$ECoG signals captured at different times,
with vertical axis (time axis) corresponding to frame number, while the other two axis represent the spatial arrangement of $\mu$ECoG electrodes; 
        }}}
\label{segments}
\end{figure}

 \begin{figure}
        \centering
        \begin{subfigure}[b]{0.12\textwidth}
                %\centering
                \includegraphics[width=\textwidth]{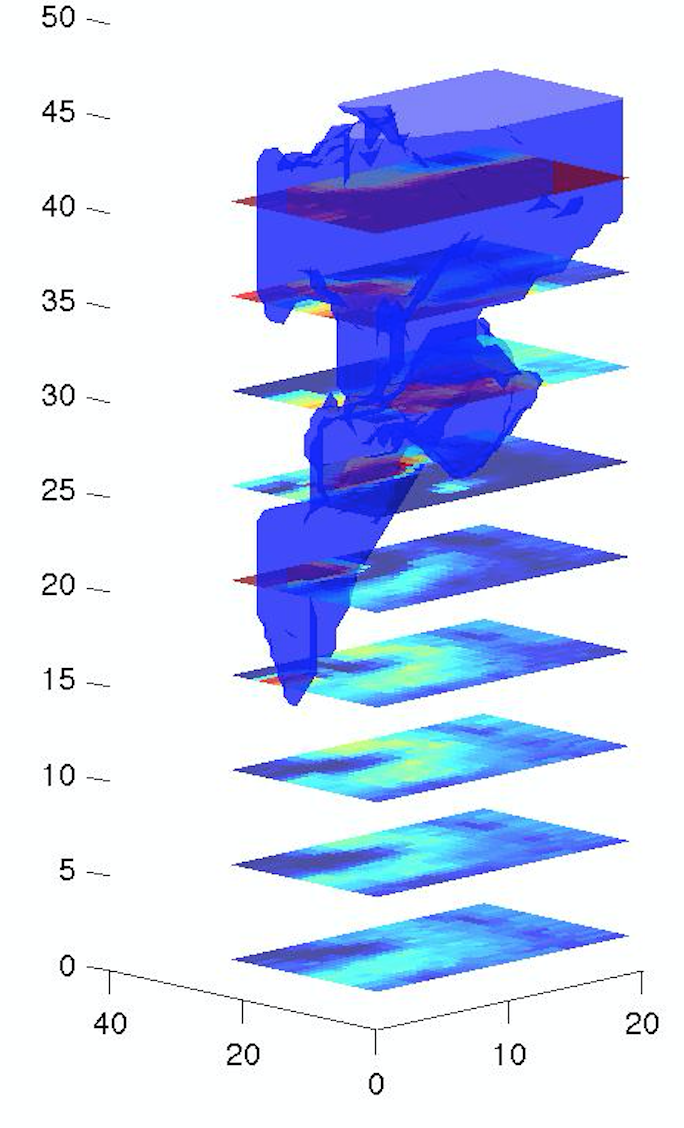}
                \caption{ }
                \label{spiral}
        \end{subfigure}%
	\begin{subfigure}[b]{0.12\textwidth}
                %\centering
                \includegraphics[width=\textwidth]{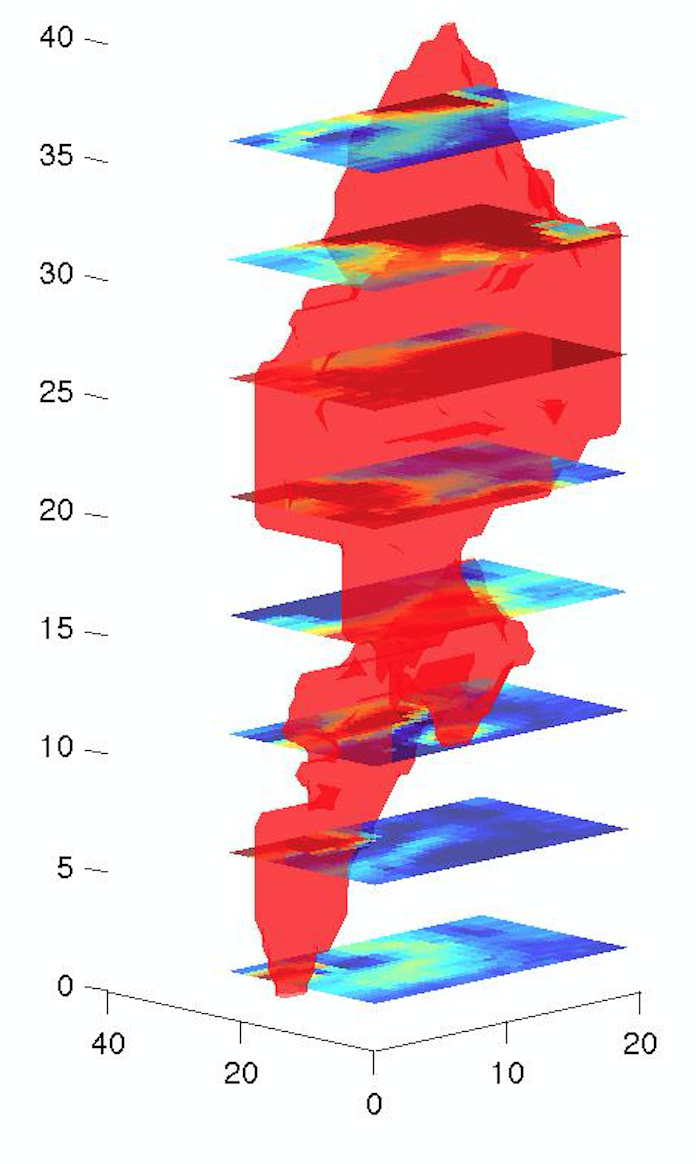}
                \caption{}
                \label{spiral traject}
        \end{subfigure}%
	%add desired spacing between images, e. g. ~, \quad, \qquad etc.
          
        \begin{subfigure}[b]{0.12\textwidth}
                %\centering
                \includegraphics[width=\textwidth]{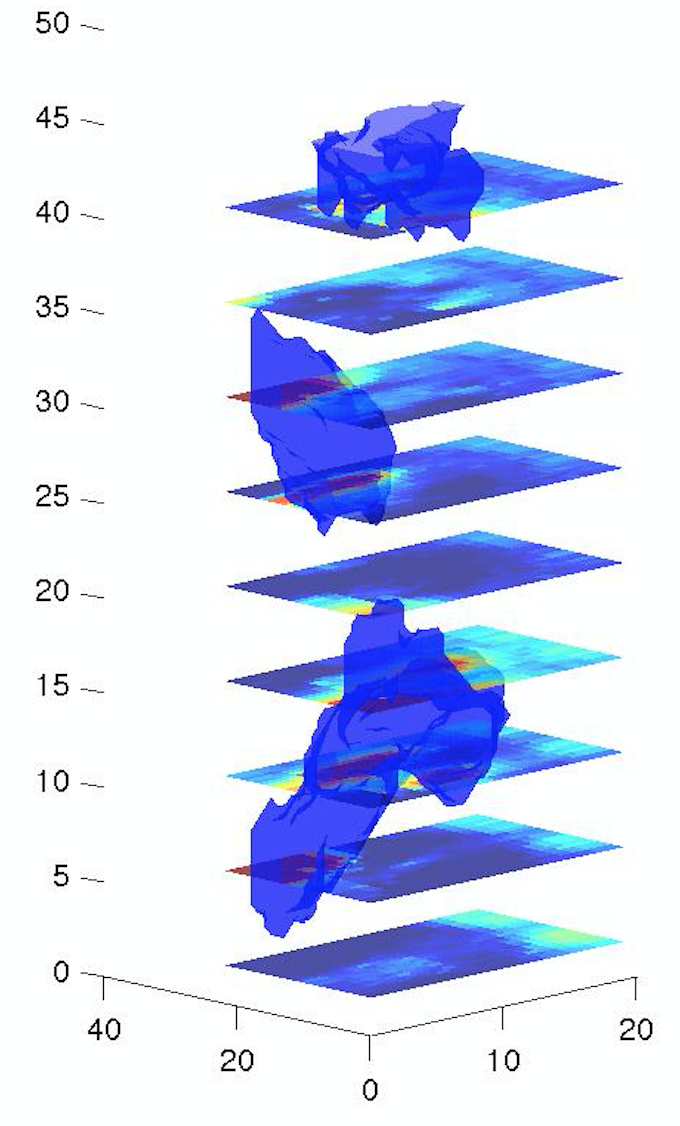}
                \caption{}
                \label{plane}
        \end{subfigure}
	\begin{subfigure}[b]{0.12\textwidth}
                %\centering
                \includegraphics[width=\textwidth]{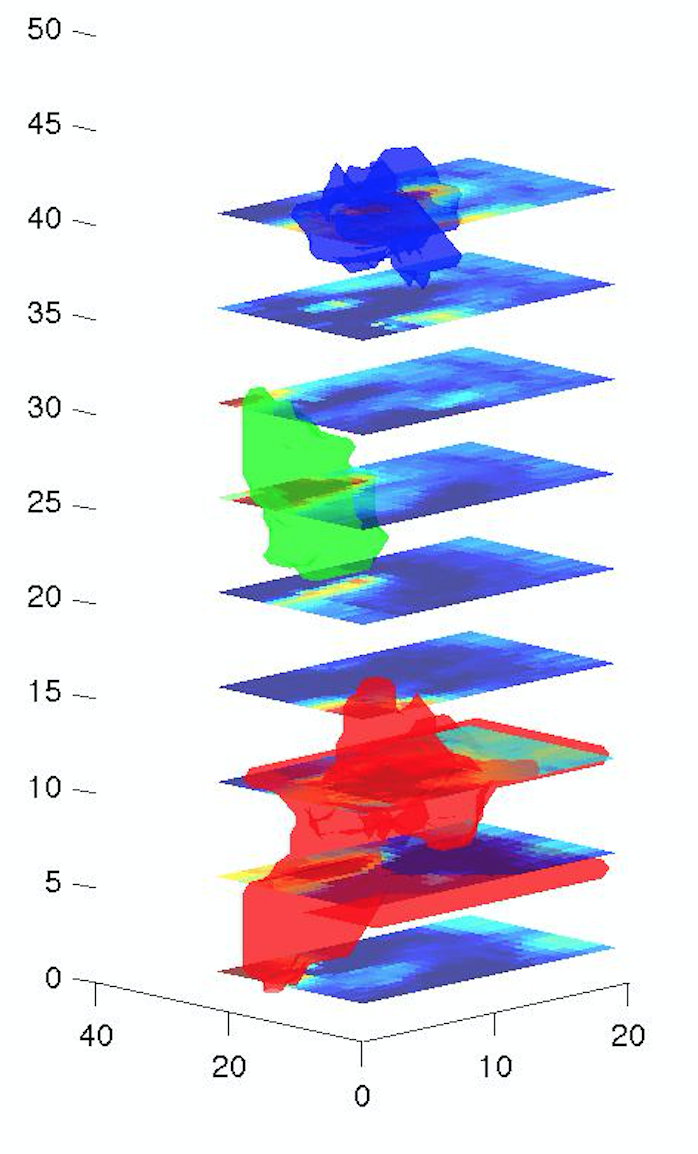}
                \caption{}
                \label{plane traject}
        \end{subfigure}
        \caption{\textit{\small{Comparison of fixed length segmentation \cite{delay_map1} (subfigure a,c) against using region-growing based variable length segmentation (subfigure b,d). With fixed length segmentation, the entire 3D cube formed by extending from a selected center frame to both past and future frames by a fixed length. The high intensity voxels in all figures are indicated by transparent color. In subfigure a,c, all voxels include within each spike segment, whereas in subfigure b,d only the high intensity voxels are included within each spike segment.  Subfigure a shows a falsely detected segment by the fixed length method, which includes a truncation on top and includes many extra voxels (mainly on the bottom) that have low intensities. The detected spike segment for the similar time duration by the proposed method is shown in subfigure b, which includes all high intensity voxels that are connected. Subfigure c shows a detected spike segment that includes three separate true spikes, which are correctly separated by the region-growing method show in subfigure d.}}}
     \label{segmentation method compare}
\end{figure}

In our acquired datasets, the multi-channel $\mu$ECoG signal has high spatial and temporal correlation, and therefore each spike could be treated as a spatial-temporally connected set of voxels that have high intensity values. A simple scheme to detect spike segment is by thresholding. However, this can separate a single spike into multiple disconnected segments, or generate isolated scattered points due to the sensor noise. In our prior work \cite{delay_map1}, a spike segment is detected when a single channel has a negative value exceeding a preset threshold (0.5 mV), and then all frames that are within a certain time window from this point both in the past (60 ms) and future (100 ms) are included in the spike. Because each spike lasts for different durations depending on its initial location and motion pattern, this simple scheme sometimes includes more than one spike or an incomplete spike in a detected segment, as shown in Fig. \ref{segmentation method compare}. This false segmentation of a spike makes our wavefront prediction within a spike almost impossible. Besides only spatially connected voxels that have high intensity signal should be included in one spike, whereas this prior method includes all pixels in the chosen duration in the detected spike. To overcome these limitations, we have applied 3D region growing, which was first reported in \cite{akyildiz2013improved}. 

%Region growing is an iterative image/video segmentation method. Region growing starts by selecting pixels (called seeds) that has intensity values above a preset threshold $P_t$. Each seed is used as an initial detected region. The algorithm then examines whether any immediate neighbors of the boundary of each previously detected region, $S_n$, also have a high intensity value, compared to a threshold that was determined based on the mean, $\eta$, and standard deviation, $\sigma$ of the pixels inside the region, with $n$ indicates the region index. Specifically, a neighboring pixel of $S_n$ with intensity $P$ would be included in $S_n$ if  $P>=\eta - \alpha \sigma$.  This process continues until no more pixels could be included in the region $S_n$.  In this way, initially disconnected regions could be merged into a single region after several time steps. At the end of the region growing process, we removed very small regions (with time span less than $T_t$ ms), and finally assigned a unique integer label to each  remaining connected 3D region. 

The region growing algorithm first detects pixels with intensity greater than $P_t$. Each such pixel becomes the seed of a region. The regions are then repeatedly expanded by including their neighboring pixels that have intensity greater than $\mu - \alpha * \sigma$, with $\mu$ and $\sigma$ being the mean and standard deviation of all current pixels included in the regions. The algorithm converges until no more pixels can be included. We applied the region growing algorithm to the data (after graph filtering) captured from an acute in vivo feline model of seizures, using the parameters  $P_t=0.5 mV$, $\alpha=0.8$ and $T_t=40ms$, with $T_t$ being the minimum span of each spike segment in time. The parameters are selected based on experiments to remove short spike segments and noise. Fig. \ref{segments} shows two detected spike segments from one dataset. Fig. \ref{segmentation method compare} compares fixed length segmentation against region growing. The result of fixed length segmentation in Fig. \ref{segmentation method compare}(a) shows an incomplete spike segment, and Fig. \ref{segmentation method compare}(c) shows multiple true spikes included within one detected spike segment. These false segmentations would make spike movement prediction impossible. The new region growing method clearly overcomes these problems as shown in Fig. \ref{segmentation method compare}(b,d) respectively.

\section{spike clustering through manifold learning} \label{spike clustering through manifold learning} 
\begin{figure*}
        \centering
        \begin{subfigure}[b]{0.35\textwidth}
                %\centering
                \includegraphics[width=\textwidth]{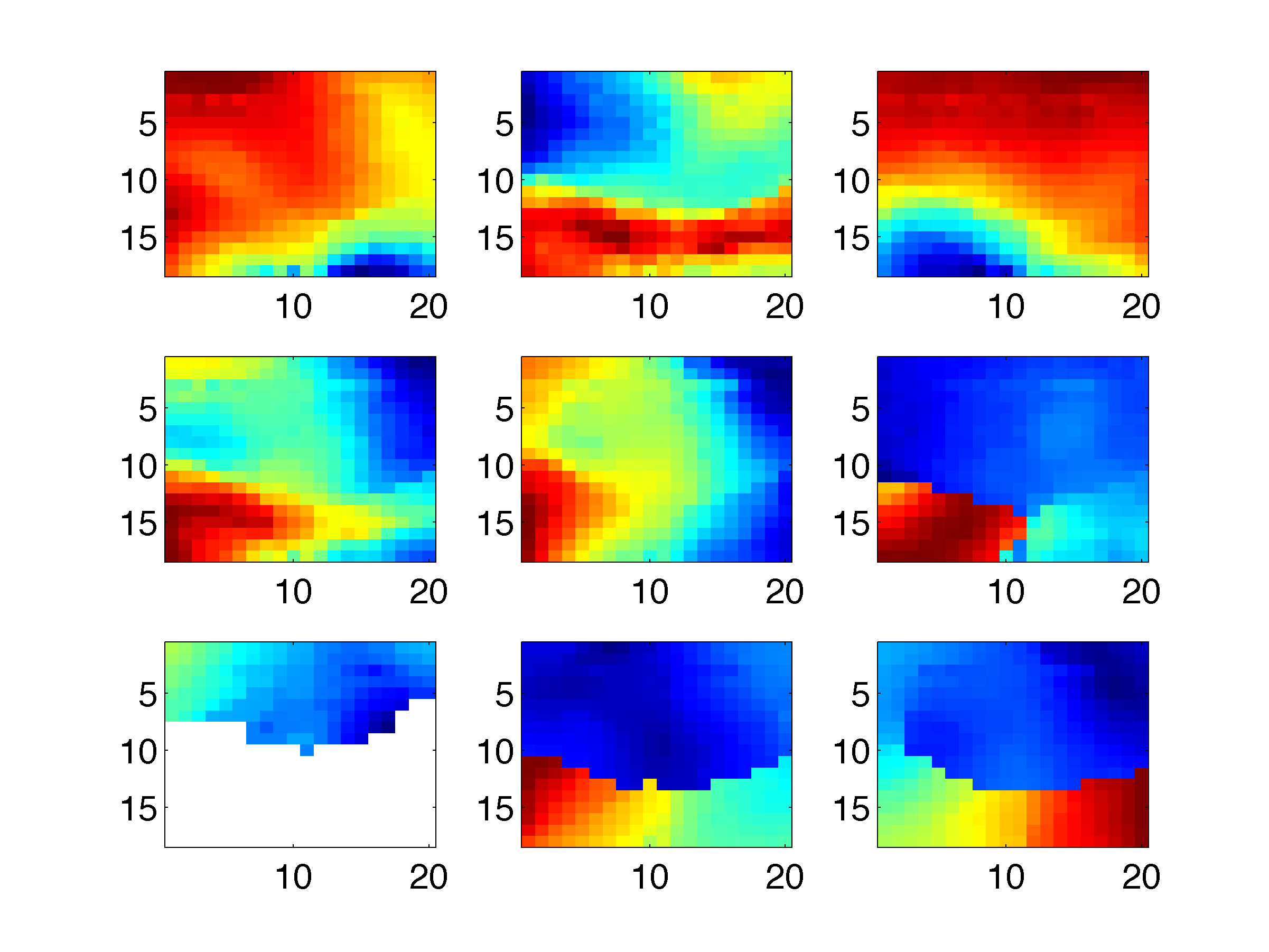}
                \caption{delay map}
                \label{fig: delay}
        \end{subfigure}%
%        \begin{subfigure}[b]{0.25\textwidth}
%                %\centering
%                \includegraphics[width=\textwidth]{test_energy}
%                \caption{energy map}
%                \label{fig: energy}
%        \end{subfigure}%	
	\begin{subfigure}[b]{0.35\textwidth}
                %\centering
                \includegraphics[width=\textwidth]{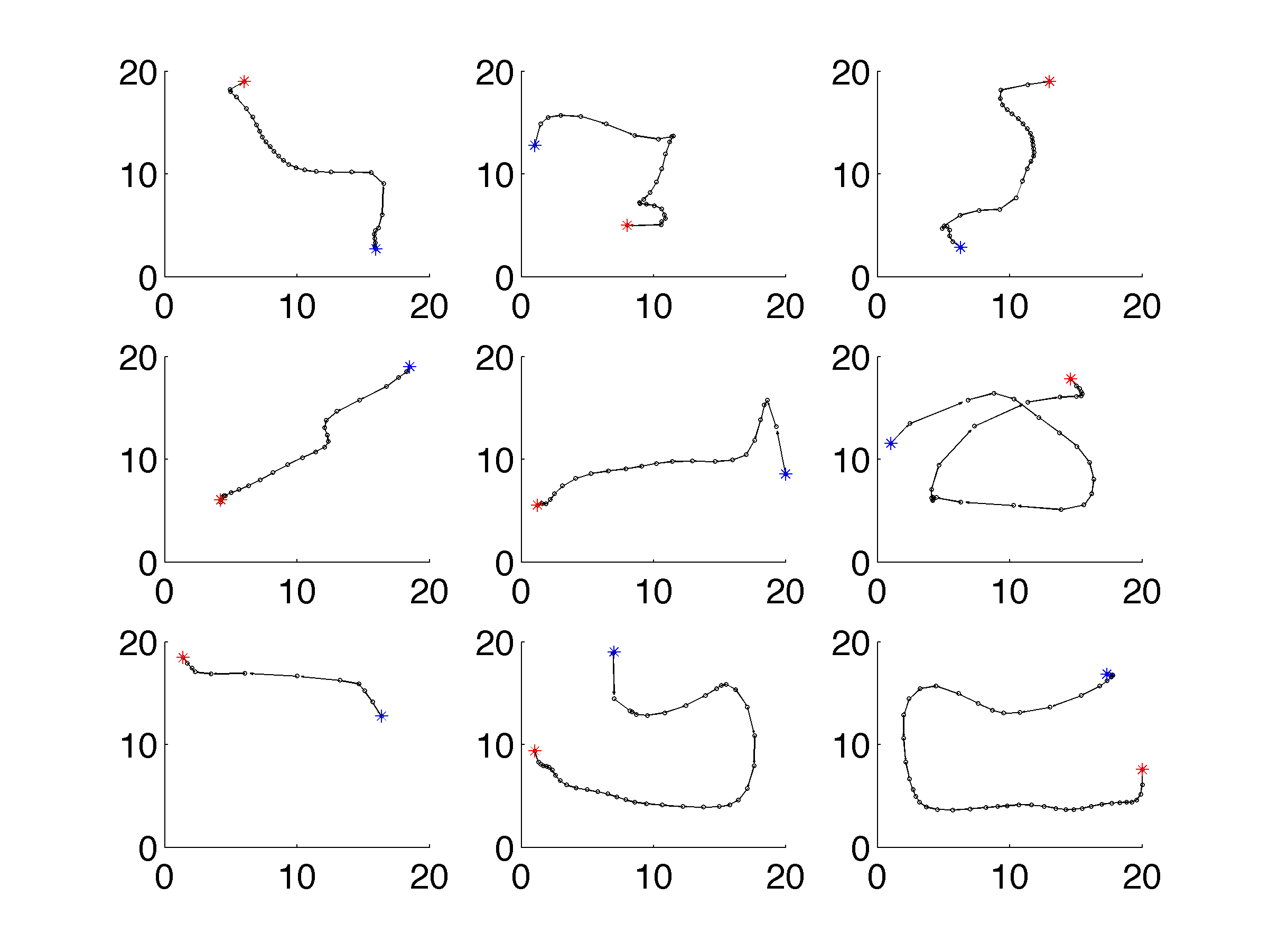}
                \caption{trajectory}
                \label{fig: delay}
        \end{subfigure}%
	 \begin{subfigure}[b]{0.35 \textwidth}
                %\centering
                \includegraphics[width=\textwidth]{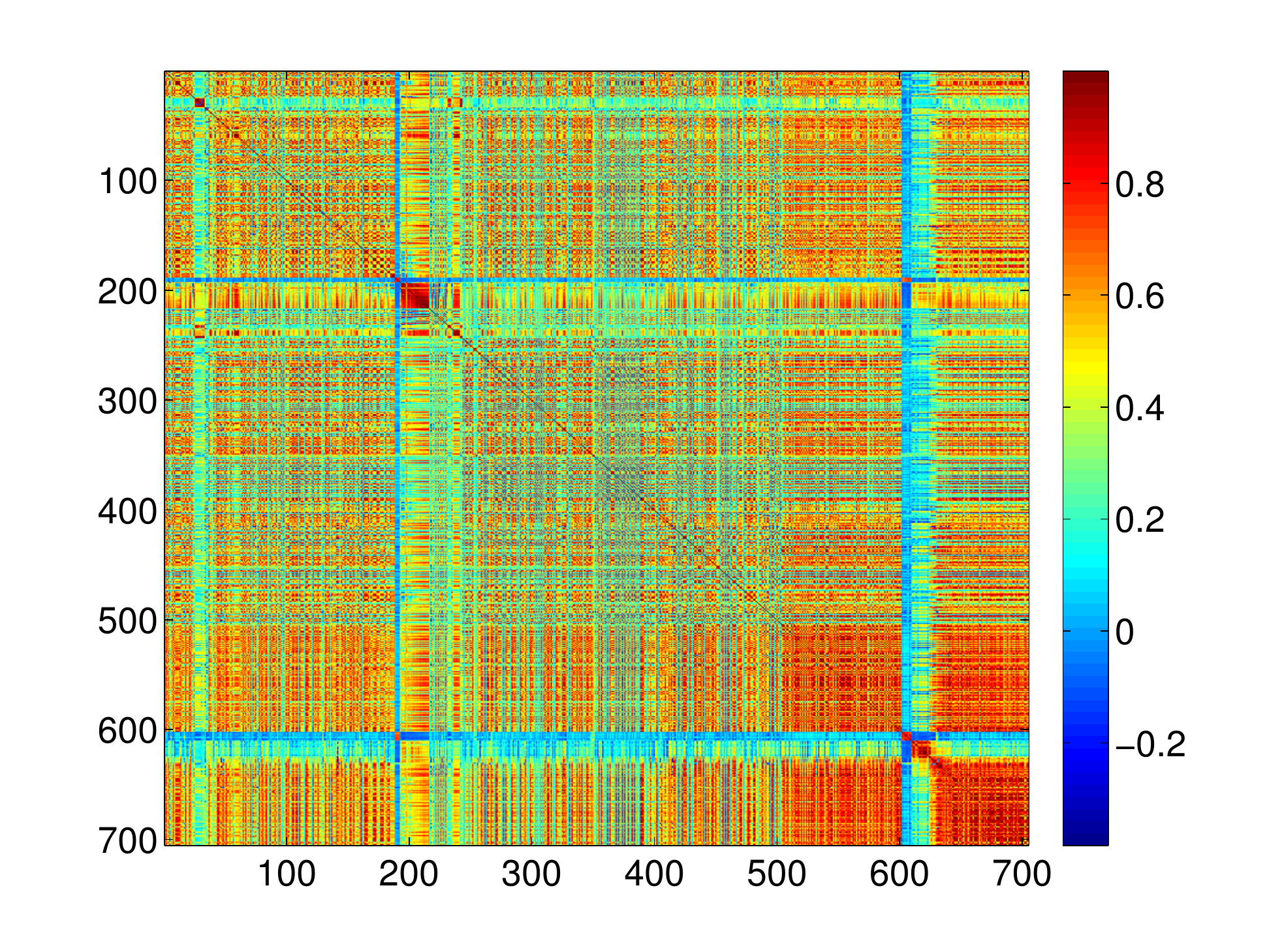}
                \caption{correlation map}
                \label{fig: delay}
        \end{subfigure}%
\caption{ \textit{\small{Examples of extracted features. Subfigures a and b are the delay and trajectory features of the 9 spikes closest to the centroids of the 9 dominant clusters identified by Isomap embedding (Section \ref{Manifold Projection}) and DPM clustering method (Section \ref{Clustering Using the Dirichlet Process Mixture Model}). 
\\In a, blue indicates areas of earliest activation while red indicates latest activation. White regions are not included in the active spike region. 
%\\b) In b, blue indicates low energy, red indicates high energy.
In b, the starting position of the wavefront is marked with a blue dot, while the ending position is marked with a red dot. The line indicates the path of the wavefront (the energy-weighted centroid).
Subfigure c is the correlation map of all 741 spike segments. Similar column vectors indicate that the corresponding spike segments had a similar propagation pattern. For example, the similarity evident in column vectors 190 to 220 could be interpreted as forming a natural cluster. }}}
\label{delay energy}
\end{figure*}

 For each of the detected spike segment, we want to characterize its spatial-temporal characteristics relying on pairwise metric that describes the relationship between spikes.  Even with the most cleverly designed feature descriptor, spikes still live in a high dimensional Euclidean space that has a non-linear lower dimensional structure. For any type of clustering algorithms to work properly, a clever way to unfold this manifold is necessary. In this section, we will first describe the pairwise metric to describe the manifold that spike segments live in. And then we present algorithms to unfold this structure and cluster patterns. 

        %\caption{Comparison of fixed length segmentation scheme against using variable length region based segmentation method. Subfigure b,d is using variable length region based segmentation method and Subfigure a,c is using fixed length segmentation method as described in \cite{delay_map1}. Subfigure a is showing a spike is falsely segmented by fixed length segmentation method, where as Subfigure c is showing spikes being falsely connected. The 3D volume in subfigure a,c are the high intensity pixels, where in fact the entire volume is considered as spike with fix length segmentation. In subfigure d, different spike segments are represented with different colors}
        
\subsection{Delay Map and Wavefront Trajectory} \label{Delay Map and Wavefront Trajectory} 
Before we delve deep into pairwise metric, we quickly describe two simple features for visualizing spike patterns. Following our prior work \cite{delay_map1}, for each spike segment we computed a delay map, which has the same dimension as the 2D sensor array, and each element indicates the delay of the signal at that sensor with respect to a reference signal. For the channel that is not in the segmented region over the entire duration of the spike segment, a 360 ms delay is assigned, same as the longest duration of spike segment in our dataset. The reference channel is the channel that has the highest sum of energy during this particular spike. Fig. \ref{delay energy}(a) shows sample delay maps for 9 representative spike segments. As can be seen the delay map is an effective and efficient way to characterize the spike motion. 

We also determined the wavefront position at each instant in time by determining the energy-weighted centroid of the spike region. Denote the set of pixels that belong to the spike region at time $t$ by $S_n(t)$, and the signal of the channel $(x,y)$ at time $t$ by $p(x,y,t)$. The $x$-position of the wavefront at time $t$ was determined by 
\begin{equation*}
\bar x_n(t) = \frac{\displaystyle\sum_{(x,y,t)\in S_n(t)} |p(x,y,t)|^2  x}{ \displaystyle\sum_{(x,y,t)\in S_n(t)} |p(x,y,t)|^2}.
\end{equation*} 
The y-position was determined similarly. Fig. \ref{delay energy}(b) shows the wavefront trajectories of the same set of 9 representative spike segments. The wavefront trajectory (a vector consists of the x and y positions from the beginning to the end frame of a spike) can also be used as the descriptor of a spike.
%9 sample of representative trajectories. And the wavefront trajectory prediction problem would be introduced in section\ref{TRAJECTORY PREDICTION}.

 \subsection {Raw Signal Correlation Map}
Since the number of spikes we got from our dataset is large, instead of coming up with hand crafted features, a better way is to let the data speak for itself. As the spike segments lives in a high dimensional space, one way of finding a low dimensional manifold for this kind of data is through a pairwise relationship description. We used the correlation of spike segments as the pairwise relational metric. Assuming that the signals corresponding to two spike segments have duration $n$ and $m$ with $n\geq m$. Denote their corresponding video signals by $X\in P^{N_1N_2n}, Y \in P^{N_1N_2m}$, where the intensity values at voxels not belonging to the spike segment are set to zero. $N_1, N_2$ represent channel dimension of the 2D sensor array. To compute the correlation between $X$ and $Y$,  we found a length-$m$ segment in $X$ that has the highest correlation coefficient with $Y$, and used this maximum as the correlation between $X$ and $Y$. The raw signal correlation matrix $W$ has a dimension of $N \times N$, $N$ is the total number of segments in the training set. The motivation for using the correlation as descriptor is that similar segments should have similar correlations with respect to other segments. Correlation may not depict the spatio-temporal characteristics of the spike segment directly, but it is crucial for unveiling the nonlinear subspace that similar spike patterns live on as we will mention next. An example raw correlation map is shown in Fig. \ref{delay energy}(c).

%The delay maps provided region shape information, by using $N_1 *  N_2$ features.
%
%\subsection{Energy Map}
%In addition to the delay maps, we calculated energy maps (Fig. \ref{delay energy} (b)) to characterize the energy distribution among all channels during a spike. For each channel its energy is determined by the mean square of signal within that spike.
%%\newcommand{\defeq}{\mathrel{:\mkern-0.25mu=}}
%E_S(x,y) \defeq \displaystyle\sum_{t: (x,y,t) \in S} {  {| p(x,y,t) |}^2
%}
%\end{equation*}

%In our application we found out by concatenating delay map and energy map together we would get better clustering performance. Before running any clustering algorithm, we pre-processed the data to normalize the mean and variance of each feature calculated over all detected spike segments. For each feature set discussed earlier or several sets combined, 
%We further tried to reduce the feature dimension in an unsupervised way to remove the intrinsic redundancy 
%within the raw features by using principle component analysis (PCA). 
%PCA is a mathematical procedure 
%that uses orthogonal transformation to convert a set of observations of possibly correlated variables into 
%a set of linearly uncorrelated variables. Thus by applying an orthogonal transform, original 
%$ x^{(i)}  \in \mathbb{R}^n $ renders a new vector $y^{(i)} \in \mathbb{R}^k$ with a lower 
%dimension k,which forms a uncorrelated representation of $x^{(i)}$. We iteratively include transformed feature space 
%with largest variance until $>99\%$ of data variance was retained.  

\subsection{Manifold Projection} \label{Manifold Projection}
Since the spike segments live on a high dimensional space, the goal of manifold learning is to construct a low dimensional embedding that maximally preserves the local structure after unfolding. Unlike its counterpart principle component analysis (PCA) \cite{PCA}, which tries to find a linear mapping, manifold embedding usually assumes nonlinear structure. In our case, the nonlinear manifold is constructed through the correlation matrix $W$ from the previous paragraph. In the following paragraphs, we briefly describe one method to unveil the manifold. This approach starts by building a graph adjacency matrix with each vertex of the graph represents a spike, each edge represents certain relationship between two spikes.

\subsubsection{Isomap}
Isomap \cite{Isomap} preserves the geodesic distance between each pair of points after unfolding the data. In our case, adjacency matrix $M$ of the dataset is constructed by using correlation matrix $W$ as: 
\begin{equation}
M_{ij} = \Bigg\{
\begin{aligned}
& (1- W_{ij})/2  & \hspace{2mm}  \text{if i,j: k-nearest neighbors}\\
 &\infty & \text{otherwise} \hspace{2mm} 
 \label{Isomap_equation}
\end{aligned}
\end{equation}

The geodesic distance matrix $D$ is defined by the shortest path between each two vertices using the adjacency matrix $M$. The normalized geodesic distance matrix $\tau(D)$ is defined as:
\begin{equation}
\begin{aligned}
& \tau(D) = -HSH/2 \\
& S_{ij} = D_{ij}^2 \\
& H_{ij} = \delta_{ij} - 1/N \\ 
 \label{Isomap_equation2}
\end{aligned}
\end{equation}
Finding the $d$ dimension embedding then can be resolved by finding the $d$ eigenvectors of matrix $\tau(D)$ with smallest eigenvalues. Given a spike segment, its manifold feature vector consists of the projections of its geodesic distance vector onto these eigenvectors. 
The detailed algorithm is described in \cite{Isomap}.

 The nearest neighborhood number $k$ in Eq. \ref{Isomap_equation} is selected to be $log(N)$ as suggested in \cite{LE_tutorial}, and the adjacency graph does not have to be fully connected. In fact by setting $k$ around $log(N)$, the adjacency graph we built has a few subgraphs which only has one single vertex. We have observed that constructing the adjacency matrix this way also serves the purpose of removing the impact of some irregular/uncommon segments that is caused by noise or false segmentation. The larger subgraphs have interesting and clean spike patterns and some of which highly correlate with seizure onset. Each large subgraph may consist of more than one spike pattern, therefore further clustering is necessary in order to separate the spike patterns within each subgraph. For our dataset, we have found that besides those subgraphs with with one node, there are a few small subgraphs (each with less than 20 nodes) plus one large subgraph. We ignore the single node subgraph, and treat each of the remaining small subgraphs as one cluster, and further separate the large subgraph into multiple clusters using manifold-based clustering. That is, we determine the eigenvectors for the large subgraph and project the correlation vectors of the spike signals in this subgraph into the eigenvectors to form their embedded features. We then apply an unsupervised clustering method on these embedded features as described below. The projection dimension $d$ in Isomap is selected by finding the largest gap between eigenvalues, and is set to 20 in our experiment.

\subsection{Clustering Using the Dirichlet Process Mixture Model} \label{Clustering Using the Dirichlet Process Mixture Model} 
To find spike patterns from low dimensional embeddings, we used Dirichlet Process Mixture (DPM) model for clustering. Unlike k-means and the Gaussian Mixture Model, the DPM model does not require that the number of clusters in the dataset is known a priori. This nonparametric Bayesian model learn the number of clusters from the dataset.
DPM is a special mixture model, where the mixing coefficient  is a random variable that follows a certain distribution. A special case of the DPM has the Gaussian distribution as its base distribution and uses the Beta distribution with a concentration
parameter to model the mixing coefficient. We used Mean-Field Variational Inference\cite{DPM} to learn the DPM parameters for a given set of samples. Once the parameters were determined, a given sample was assigned to the mixture with which it has the highest likelihood. 

%Specifically, this type of DPM is defined by:
%
%
%\begin{equation}
%  p(x) = \displaystyle\sum_{k=1}^K \pi_k \mathcal{N}_{\mu_k, \sigma_k}(x)
%\end{equation}
%where 
%$$ \pi_k = \beta_k \prod_{l=1}^{k-1} (1-\beta_l)\\$$
%$$ \beta_l \sim 
%B(1, \alpha) $$
%$B(1,\alpha)$ determines the beta distribution. The concentration parameter $\alpha$ determines the structure of the cluster. As it increases, the mixing weights $\pi$ become more uniform, which favors even clusters. As $\alpha$ decreases, the process favors uneven clusters, so that most samples concentrate in a few large clusters. One inherent problem with the K-means clustering method is that it tends to generate clusters with similar sizes, even if the natural clusters have variable sizes. By selecting the correct $\alpha$, DPM based clustering could be particularly effective when the training data tend to form clusters that vary greatly in size, which is the case of our data. We used Mean-Field Variational Inference\cite{DPM} to learn the DPM parameters for a given set of samples. Once the parameters were determined, a given sample was assigned to the mixture with which it has the highest likelihood. 
%With this algorithm, only an upper bound needs to be provided for maximum possible number of mixtures. When the number of natural clusters in the data is smaller than the upper bound, only some clusters have non-zero mixture coefficients for a suitable concentration parameter $\alpha$. 

\subsection{Clustering Results Comparison}
We applied our manifold clustering techniques onto 741 spike segments obtained from in-vivo recordings of a feline model of epilepsy. 
To compare with prior work \cite{delay_map1,Ann_new_delay}, we used the same portion of dataset 1 (Table \ref{dataset}). A detailed description of the dataset is given in Section V.    

\begin{figure*}
	%add desired spacing between images, e. g. ~, \quad, \qquad etc.
        \begin{subfigure}[b]{0.35\textwidth}
                %\centering
                \includegraphics[width=\textwidth]{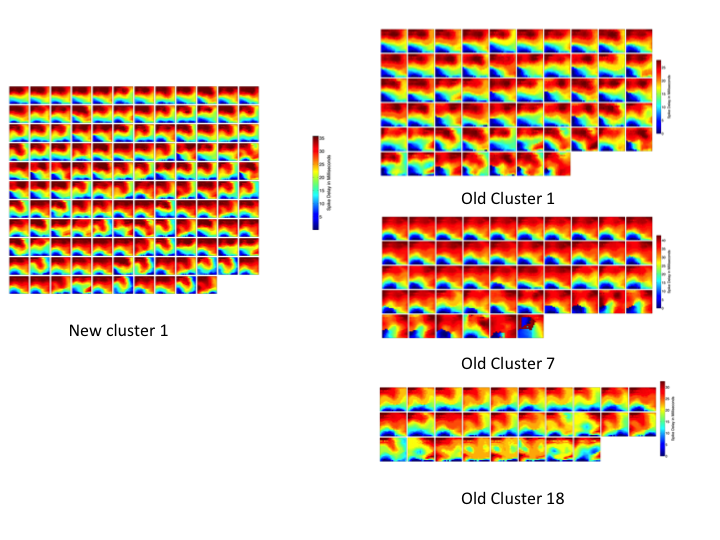}
                \caption{}
                \label{}
        \end{subfigure}
	\begin{subfigure}[b]{0.35\textwidth}
                %\centering
                \includegraphics[width=\textwidth]{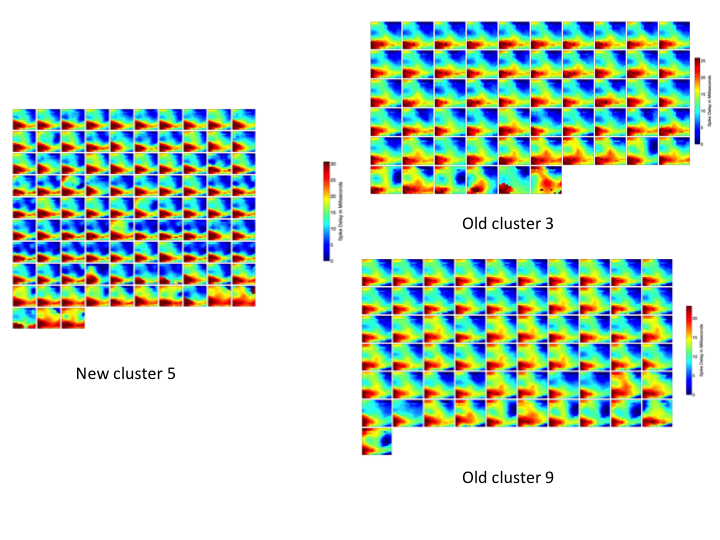}
                \caption{}
                \label{}
        \end{subfigure}%
        \begin{subfigure}[b]{0.35\textwidth}
                %\centering
                \includegraphics[width=\textwidth]{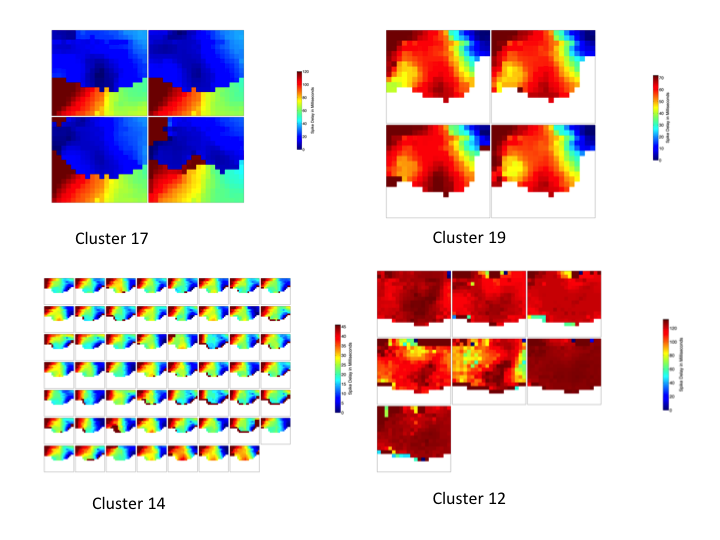}
                \caption{}
                \label{}
        \end{subfigure}%
\caption{\textit{\small{Cluster result comparison with \cite{delay_map1}. Subfigure a,b demonstrate two cases where several clusters in \cite{delay_map1} should be merged into one. Subfigure c displays several unique clusters that \cite{delay_map1} failed to discover. Note that the spike patterns in subfigure c are very rare in the datasets and tend to be missed with standard linear embedding and K-means clustering method.
}}}
\label{cluster compare}
\end{figure*}

We have found that DPM on isomap projections have derived cleaner spike patterns within each cluster compared with the clusters derived in \cite{delay_map1}. 
In this prior work, we first use fixed length segmentation to find spikes and then used delay and energy as our feature representation. PCA is applied on concatenated feature maps to find a linear low dimensional embedding and then adopt K-means algorithm for spike clustering. The region based spike segmentation in this work could adaptively find neural activity restricted to a local region and allow variable length of spike segments. This ensures cleaner spike segmentation. Furthermore, the number of raw pixels belonging to each spike segment generally exceed tens of thousands, and it is highly likely that these spike segments lie on a non-linear low dimensional manifold than a linear one. PCA as a linear dimension-reduction method used in \cite{delay_map1} fails to capture the non-linearity in the dataset. 
One limitation of K-means is that it requires the knowledge of the number of clusters, which is hard to determine in the case of spike clustering, even with human inspection.
One observation we have is that some types of spikes are much more frequently observed than others. K-means clustering usually works well only when the clusters have comparable sizes. On the other hand, DPM, as a special mixture model, does not require the knowledge of the number of clusters a priori and can generate unevenly distributed clusters with the proper selection of the concentration parameters. We present visualization results of spike pattern clusters in Fig. \ref{cluster compare}. Figure \ref{cluster compare}(a) and (b) demonstrate two cases where clusters derived in \cite{delay_map1} should be merged into one. Fig. \ref{cluster compare}(c) shows new clusters that \cite{delay_map1} failed to discover.

\section{wavefront trajectory prediction}\label{TRAJECTORY PREDICTION}
As described earlier, the wavefront at any particular frame time is obtained by averaging the locations of the channels weighted by the channel signal magnitude. Because the signal of the leading edge of the spike usually has much higher intensity compared to the rest of the spike, this serves as an approximation of the wavefront. Let $(x_t, y_t)$ denotes the wavefront location at frame $t$. 
The problem of wavefront prediction is to predict future locations of the wavefront at frames $t>k$, given their locations in $k$  beginning frames $t=1,2, \ldots,k$. One way to predict this time series is by polynomial regression, where we model the trajectory as a degree-$n$ polynomial function of $t$:
\begin{equation}
\begin{aligned}
x_t &= a_{n}t^n + a_{n-1}t^{n-1}+ \cdots + a_0 \\
y_t &= b_{n}t^n + b_{n-1}t^{n-1}+ \cdots + b_0 \\
\end{aligned}
\end{equation}
Because $x_t$ and $y_t$ are individually parameterized by polynomial function, without loss of generality, we give the mathematic formulation of solving $a_n,\cdots, a_0$ as the polynomial regression problem in Eq. \ref{trajectory_fit_math}. The parameters of this polynomial function are regressed from the first $k, k \leq n$ observations of the trajectory $x_t, t = 1,\cdots, k$ and then we use the coefficients found to generate wavefront positions in the future. Specifically, we determine the coefficients by solving the following problem:
%Once a patient is already in seizure stage, we need to predict the wavefront patten or trajectory of the wavefront in the near future for seizure suppression. Predicting wavefront pattern could be generalized to either prediction wavefront trajectory or the raw signal. One point in wavefront trajectory would be the weighted signal in one particular frame, and the trajectory would be the time series of such points. Then predicting of the trajectory of one spike given a couple of observation at the beginning of the trajectory is solving a curve fitting problem, where we predict trajectory's location $p(x_t,y_t)$ as a polynomial function of $t$. For prediction of wavefront where we can only use a small portion of trajectory to estimate the future of trajectory, it would yield unreliable result as shown in Fig. \ref{trajectory_fit}. The mathematical formulation of the problem would be Eq. \ref{trajectory_fit_math}, with $x$ being the observed position, $a$ being the polynomial coefficients.

\begin{equation}
\begin{aligned}
 &\underset{a}{\text{minimize}} \norm{x - T a}^2 \\
where \hspace{0.1cm} &x= [x_1,x_2,...,x_k]^T \\
&a= [a_0,a_1 ,...a_{n}] ^T\\
&T =  
\begin{pmatrix}
1&1& \cdots & 1 \\
1&2& \cdots & 2^n \\
1&3& \cdots & 3^n \\
\vdots & \vdots &\ddots & \vdots \\
1 & k &\cdots & k^n 
\end{pmatrix}
\end{aligned}
\label{trajectory_fit_math}
\end{equation}

%\begin{equation*}
%
%\end{equation*}

\begin{figure}
\centering 
\begin{subfigure}[b]{1\textwidth}
	\includegraphics[width = 0.45\textwidth] {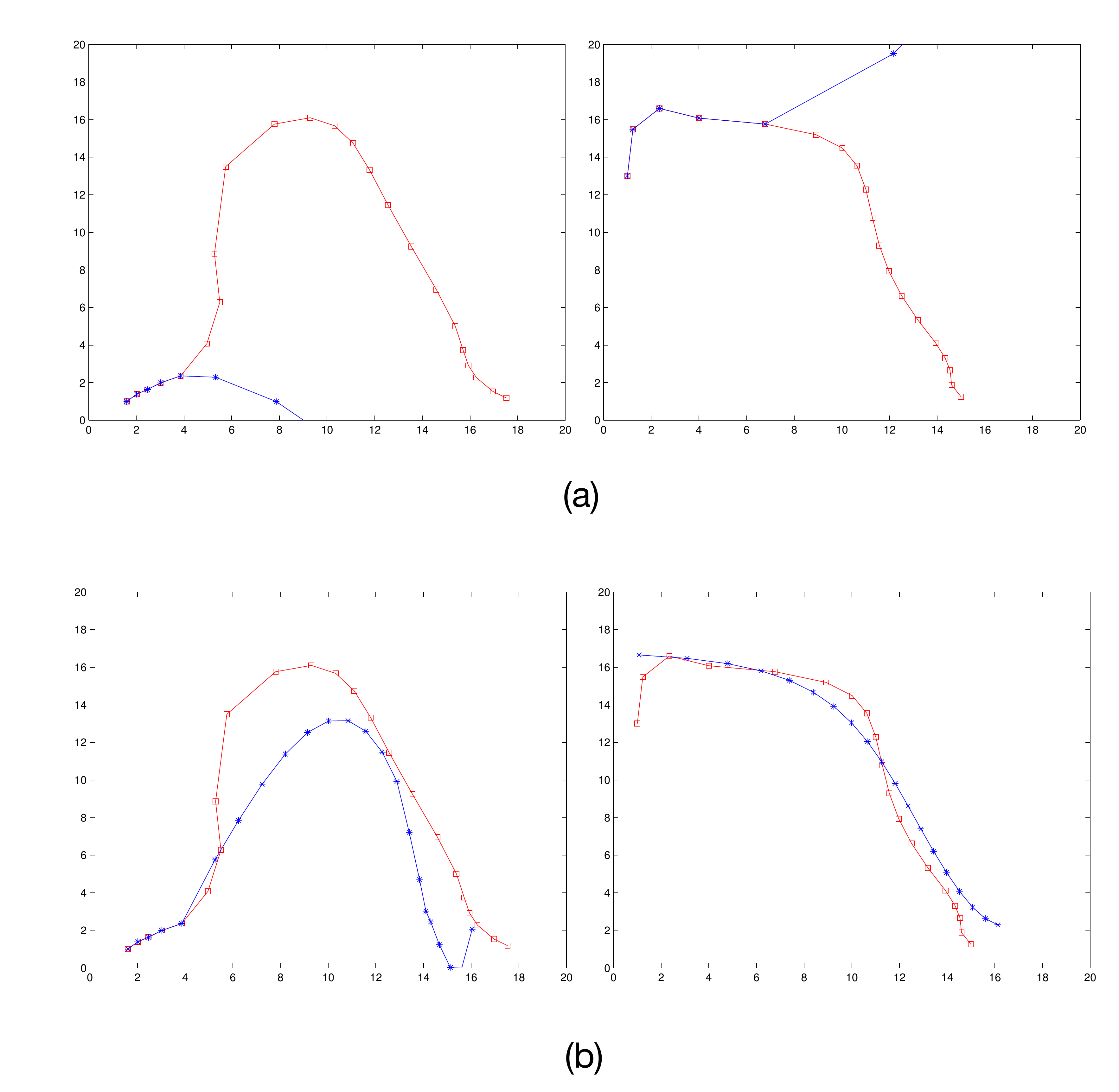}
	\label{}
\end{subfigure}
%~
%\begin{subfigure}[b]{0.3\textwidth}
%	\includegraphics[width = \textwidth] {plots/fit_use_5_2}
%	\label{}
%\end{subfigure}
\caption[Trajectory prediction with polynomial regression and mean-based regularization on coefficients of cluster]{\textit{\small{Trajectory prediction with polynomial regression and mean-based regularization on coefficients. The spike trajectories are predicted given five observations by fitting 4 degree polynomial function. Red curve is the ground truth, blue curve is the prediction results. Subfigure (a) shows the results using polynomial regression. The overlapped 5 points of prediction and ground truth is the first five observation of that particular trajectory.
Subfigure (b) demonstrates the predictions by further adding mean-based regularization.}}}
\label{trajectory_fit}
\end{figure}
%
%
%\begin{figure}
%        \centering
%        \begin{subfigure}[b]{0.22\textwidth}
%                %\centering
%                \includegraphics[width=\textwidth]{plots/2only}
%                \caption{}
%                \label{}
%        \end{subfigure}%
%%        \begin{subfigure}[b]{0.22\textwidth}
%%                %\centering
%%                \includegraphics[width=\textwidth]{plots/2only}
%%                \caption{}
%%                \label{}
%%        \end{subfigure}%
%  ~
%        \begin{subfigure}[b]{0.22\textwidth}
%                   %     \centering
%                \includegraphics[width=\textwidth]{plots/17only}
%                \caption{}
%                \label{13}
%        \end{subfigure}%
%%        \begin{subfigure}[b]{0.22\textwidth}
%%                     %   \centering
%%                \includegraphics[width=\textwidth]{plots/17only}
%%	     \caption{}
%%                \label{}
%%        \end{subfigure}%
%\caption{\textit{\small{Trajectory prediction results using mean-based regularization on coefficients of cluster mean using Eq.\ref{opt_setup}. Red curve are ground truth trajectories and blue curve are prediction results given the first five points as observation. }}}
%\label{prediction_plot_with_reg}
%\end{figure} 

The least square solution is $a = (T^T T)^{-1}T x$. As we can only use a small initial portion of the trajectory to estimate the future trajectory, it would yield unreliable result as shown in Fig. \ref{trajectory_fit}(a).

\subsection{Trajectory Prediction} \label{Trajectory Prediction with Mean-based Regularization} 

As shown before, predicting the wavefront trajectory of a spike from a few samples at the beginning of the trajectory is hard, as the wavefront trajectory varies from one spike to another in a highly non-linear way. This is an extremely challenging task, without any prior knowledge of the spike pattern. By recognizing the cluster that the current spike belongs to, we can make prediction much easier. 
%Using the clustering techniques we described in previous section, we fit the polynomial coefficients with regularization.  

\begin{figure}
\centering 
	\begin{subfigure}[b]{0.4\textwidth}
	\includegraphics[width = \textwidth] {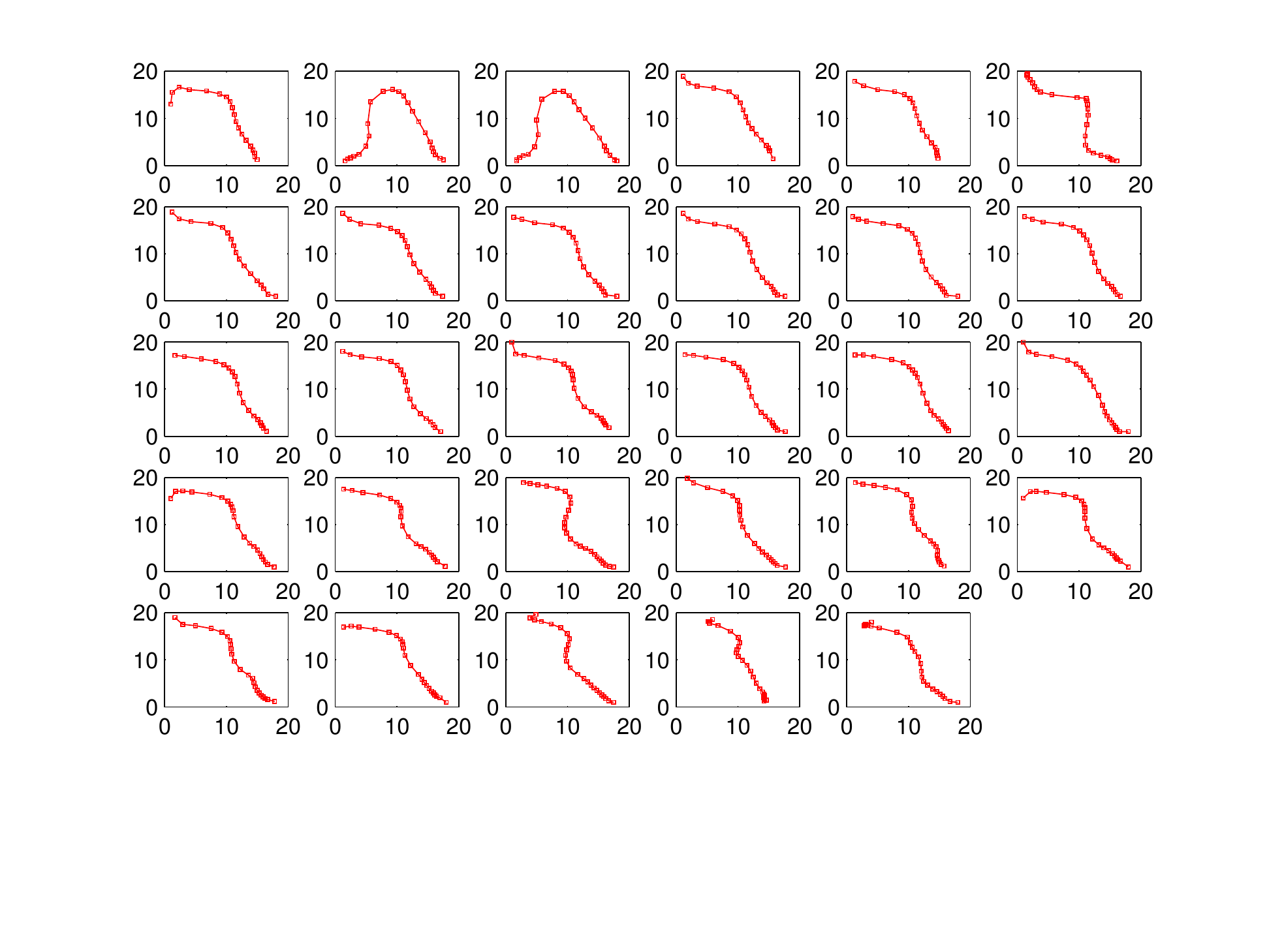}
	\end{subfigure}
\caption{\textit{\small{Trajectories of spikes belonging to the same cluster. Each subfigure is a trajectory from one particular spike segment. Clustering is performed using Isomap and DPM. As can be seen the trajectories within this cluster has a highly consistent pattern.}}}
\label{cluster_trajectory}
\end{figure}

As shown in Fig. \ref{cluster_trajectory}, the spikes in the same cluster have consistent trajectory patterns. If we know the cluster that a spike belongs to, and we also know the mean of the polynomial coefficients of the spike trajectories in this cluster based on training data, we can modify the previous least squares fitting problem with an additional constraint. Given the cluster index $i$ of a spike, we determine the polynomial coefficients using the beginning of the trajectory, with additional constraint that the coefficients should be close to the coefficient mean of that particular cluster. Let $\bar{a}$ denotes the mean of the coefficients and $Q$ is the inverse covariance matrix of $a \in \text{cluster i}$, $\norm{a}^2_Q$ stands for the weighted norm $a^T Q a$. This formulation is equivalent to find the maximum a posterior estimation of $a$ by assuming a gaussian prior distribution. 
\begin{equation}
\underset{a}{\text{minimize}} \norm{x - T a}^2 +  \norm{a - \bar{a}}^2_Q
\label{opt_setup}
\end{equation}
The optimization problem has a closed form solution.
\begin{equation*}
a^* = \bar{a} + (T^TT+Q)^{-1}(T^T(x-T\bar{a}))
\end{equation*}

Fig. \ref{trajectory_fit}(b) shows two of the prediction results using mean-based regularization. In addition to using mean-based regularization, we also compared merely using mean of the polynomial coefficients of the cluster as a strong benchmark. The root mean square error between the predicted wavefront locations and the actual locations using regularization, mean and  simple least squares fitting are summarized in Table \ref{Fitting_comparison}. To make the benchmark method of only using polynomial fitting stronger, we further restrict all predicted positions of x,y coordinates using different methods to be within the grid size. As shown in Table \ref{Fitting_comparison} prediction of mean-based regularization is within 4 pixels of the true location, on average, which is much better than prediction using polynomial fitting just based on the first 5 observations.

Of course, to use the proposed cluster-based method, one must be able to determine the cluster that the current spike belongs to. This is difficult to do using only the first five frames of a spike. We plan to investigate how to predict spike patten label based on the spike label variations in history as future work.

\begin{table}
\centering
\caption{Trajectory Prediction Error (pixel)}
\scalebox{0.8}{
    \begin{tabular}{  p{1.35cm} |  p{3cm} p{2cm} p{2cm}}
   \toprule Dataset & Cluster-based polynomial fitting with mean-based regularization & Prediction using cluster mean  & Polynomial fitting \\
    \toprule
    Dataset 1 & 7.62  & 8.10 & 20.21\\ 
    Dataset 2 & 8.31 &9.07 &23.52 \\ 
    Dataset 3 &8.13  &8.49 &21.30\\ 
   \bottomrule
    \end{tabular}
    }
    \begin{tablenotes}
      \small
      \item \textit{\small{Root mean square error of trajectory prediction for all detected spikes in different datasets. Wavefront locations in all remaining frames of a spike segment are predicted from the polynomial coefficients derived from the wavefront locations in the first five frames.}}
    \end{tablenotes}
    
    \label{Fitting_comparison}    
\end{table}
%By constraining the trajectory, we are able to get better prediction result and that is motivating ensemble learning in the next section.

%\begin{figure}
%\centering
%	\begin{subfigure}[b]{0.5\textwidth}
%	\includegraphics[width =0.25 \textwidth] {plots/2}
%	\caption{}
%	\label{}
%	\end{subfigure}
%        \begin{subfigure}[b]{0.5\textwidth}
%        	\includegraphics[width = 0.25\textwidth] {plots/2only}
%        	\caption{}
%        	\label{}
%        \end{subfigure}
%        \begin{subfigure}[b]{0.5\textwidth}
%        	\includegraphics[width =0.25 \textwidth] {plots/17}
%        	\caption{}
%        	\label{}
%        \end{subfigure}
%        \begin{subfigure}[b]{0.5\textwidth}
%        	\includegraphics[width = 0.25\textwidth] {plots/17only}
%        	\caption{}
%        	\label{}
%        \end{subfigure}
%	\caption[Trajectory prediction results comparison]{ \textbf{Trajectory prediction results comparison.} Red curve are ground truth trajectories and blue curve are prediction results using different methods. subfigure a,c are using the mean of the trajectories' cluster as prediction, whereas subfigure b, d are prediction result using Eq.\ref{opt_setup}. }
%\label{prediction_plot_with_reg}
%\end{figure}

\section{Seizure Detection and Prediction} \label{Seizure Detection and Prediction} 
\begin{figure}
	%add desired spacing between images, e. g. ~, \quad, \qquad etc.
                \centering
                \includegraphics[width=0.4\textwidth]{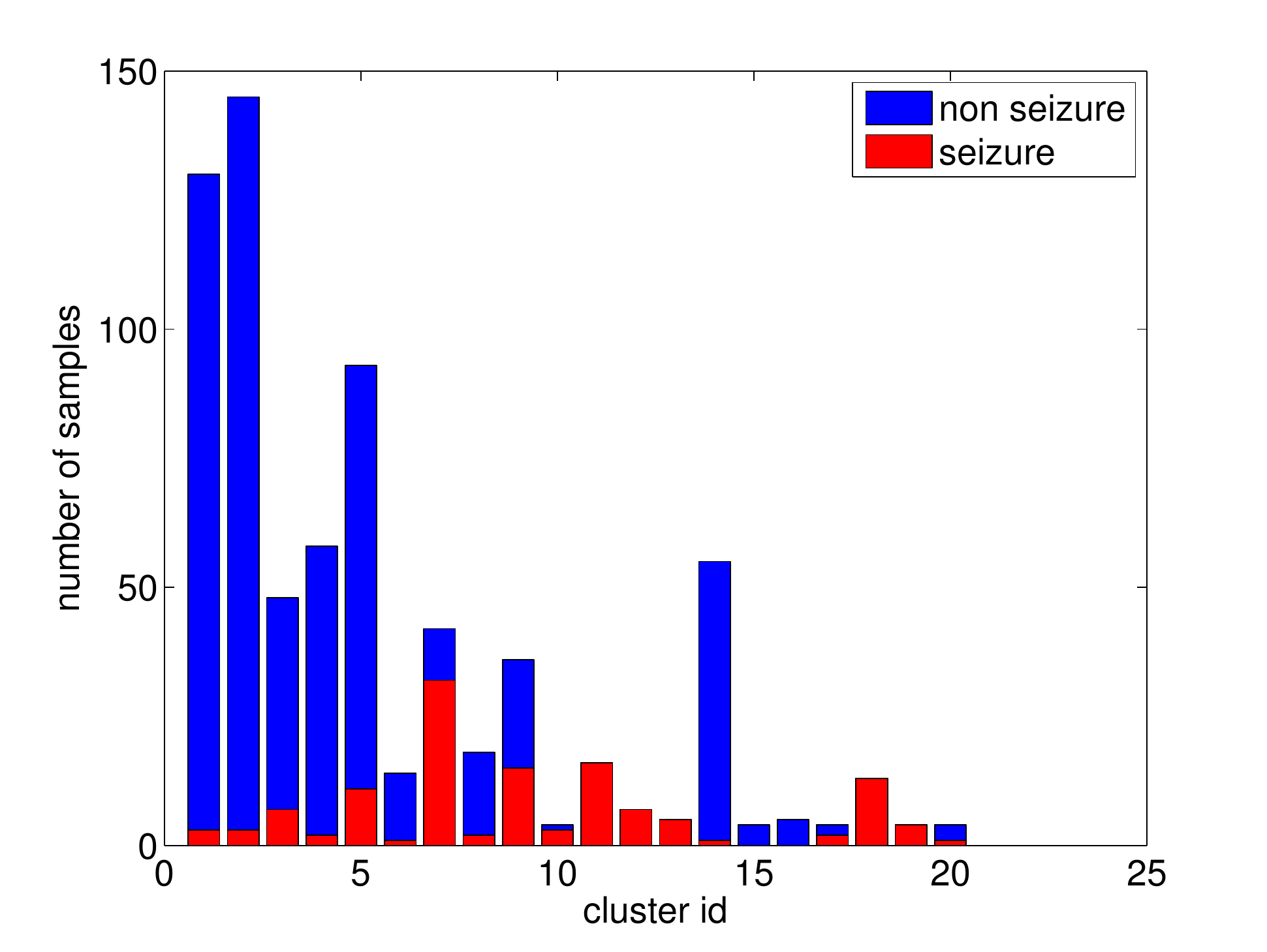}
        \caption{\textit{\small{Bar chart representing distribution of 20 identified clusters by applying DPM on Isomap projections. Using majority vote of spike label (seizure, none-seizure) of each cluster for seizure detection would render seizure detection sensitivity of: 0.5703, seizure detection specificity of: 0.9983 respectively.}}}
	\label{cluster_bar_chart}
\end{figure}

Aside from wavefront prediction, we also investigated how to detect and predict seizure onset based on the spike pattern label variation. Note that seizure prediction relying on one spike segment is unreliable. Figure \ref{cluster_bar_chart} displays the distribution of nonseizure vs. seizure spikes over the clusters found by the DPM clustering method on isomap projections. Each bar represents one of the 20 different spike patterns identified. It can be seen that ictal (during a seizure) and inter-ical (between seizure) spikes are well separated. In particular, cluster 7, 11, 12, 13, 18 and 19 mainly contain spikes that occur during seizures whereas cluster 1, 2, 3, 4, 5, 6, 14, 15, 16 are mostly spikes that occur between seizures. This result reveals a cleaner separation between seizure cluster and non-seizure clusters against \cite{Ann_new_delay}. But there are certain clusters like cluster 9 that mixes seizure patterns and non-seizure patterns. Therefore for a more robust seizure detection and prediction one has to take into account of spike pattern variation in a longer time horizon.

 In order to accomplish the goal of seizure detection and prediction, we developed a method to classify a current spike into one of the three spike categories : interictal (non-seizure), pre-ictal (pre-seizure) and ictal (i.e seizure), based on the temporal variation of spike pattern labels surrounding the current spike as illustrated in Fig. \ref{Class_struc}.
The datasets we have are from an acute in vivo feline model of seizures. There are many seizures in the dataset and the intervals between consecutive seizures are fairly short, 41 seconds on average. Therefore we labeled all spike segments that occurred up to 8 seconds before a seizure onset as pre-ictal. We consider correct identification of the pre-seizure state as seizure prediction, and correct identification of seizure as seizure detection.
For neural datasets, a significant portion of the data recording is low amplitude signal, for which we assigned a unique cluster label 0. For resting period $d$ longer than 100 ms, the number of zero labels is $ceil(d/100)$, since the average duration for spike segments is 100 ms. 
Spike segmentation was accomplished by region growing as described in Sec.\ref{spike segmentation}. To determine which cluster each spike belongs to, we first find the correlation vector of this spike with respect to all spikes in the training set, and then convert this correlation vector to a distance vector following the same procedure in computing the distance in Eq. \ref{Isomap_equation}. For each training sample which is not among the $k$ nearest neighbors of the current spike, we replace the infinity distance by the geodesic distance of the current spike to that sample through  all training samples. Then we projected the resulting geodesic distance vector onto pretrained Isomap eigenvectors to obtain manifold features. Next we evaluated the likelihood of the resulting manifold features with each mixture model in the pre-trained DPM for clustering and identified the cluster that has the highest likelihood.

\subsection{Using Multiple Hidden Markov Models Plus a Seizure Class Classifier (HMM+SVM)} 

\begin{figure*}
        \centering
%        \begin{subfigure}[b]{0.45\textwidth}
                %\centering
                \includegraphics[width =0.87\textwidth]{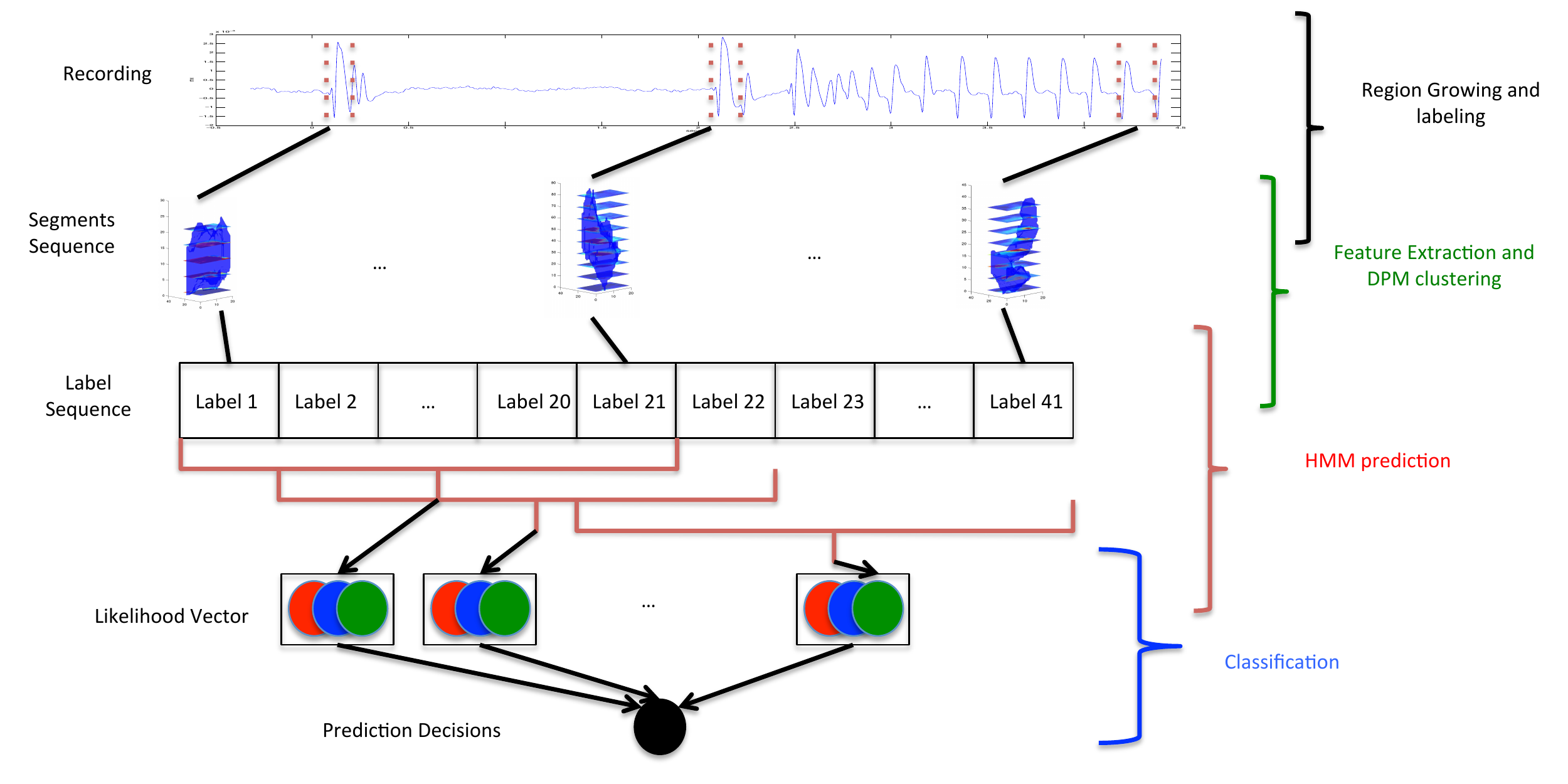}
                \label{}
%        \end{subfigure}%

\caption{\textit{\small{Illustration of the HMM+SVM seizure detection/ prediction scheme. Input of the diagram is the raw $\mu$ECoG data.  Black bracket indicates the process of segmenting the raw data into spike segments. Green bracket includes the procedure of extracting features for each spike and classifying it into one of predetermined spike clusters using pre-trained DPM model. Red bracket covers the process of multiple HMM modeling for each overlapping sequence of 21 spike cluster labels. Three different colors in each box represent the likelihood of the sequence belonging to one out of three HMMs. Blue bracket represents SVM classification of the likelihood vector computed from a total of 21 overlapping sequences centered around the current spike. The output of the diagram is the classified seizure stage for each spike. 
Seizure onset is detected or predicted when the current and previous 4 spikes  are all classified as either pre-ictal or ictal spikes.}}}
%\caption{Overall Flow structure of Learning. Detailed description of prediction structure. The current prediction is for label 21. Input of the diagram would be spike segments detected from raw-$\mu EcoG$ data. Green bracket generalizes the procedure of feature extraction followed by DPM clustering. Red bracket generalize the sliding window approach of multiple-HMM prediction algorithm. Three different colors in each box represent the likelihood of the sequence belonging to one out of three HMMs. Blue bracket represents SVM classification and the output of the diagram would be one out of three seizure stages.}
\label{Class_struc}
\end{figure*}

In this approach, we built a separate HMM for each individual seizure stage (i.e., interictal, pre-ictal, and ictal).  Let $\lambda= (A,B,\pi)$ denote the model parameters for each stage. $A_{ij}$ is the state transition probabilities, $B_{jk}$ is the emission probabilities of state $j$, and $\pi_i$ is the stationary probability of state $i$.
\begin{equation*}
\begin{aligned}
A_{i,j} &= P(q_t = S_i|q_{t-1} = S_j) \\
B_{jk}&= P(O_t = k | q_t = S_j) \\
\pi_i &= P(q_1 = S_i) 
\end{aligned}
\end{equation*}

During training, the goal is to adjust model parameters $\lambda = (A,B,\pi)$ to maximize $P(O|\lambda)$ using observation sequences in the training data for each stage. Model parameters are updated iteratively through Baum-Welch algorithm \cite{HMM}. Since we had no prior knowledge of how many states to choose, the number of hidden states was chosen to maximize $P(O|\lambda)$ through trial and error. We found that four hidden states work well for all three different HMMs. We have to emphasize that each HMM is optimized using the training data from corresponding seizure stage (i.e., interictal, pre-ictal, and ictal).

We have found that using the HMM trained separately for each seizure stage cannot accurately classify a spike during the transition period from non-seizure to seizure (or vice versa). We suspect that this is because, during the transition period, the observation sequence contains spikes from both the non-seizure period and the  seizure period. If we record the actual likelihood values for all three seizure stages in time, it is likely that, during the transition period, the likelihood for the non-seizure stage decreases, and the likelihood for the seizure stage increases. To exploit the likelihood variation pattern over time, we adopted a sliding window approach and collected the likelihood vector (consisting of the likelihood values of the three seizure stages) over each sliding window. As shown in Fig. \ref{Class_struc}, to classify the observation $O_t$ as indicated by the red double dashed vertical lines in the middle, we collected a total of $2n+1$ observations. For each sliding window $k$ containing $n$ observations, we derived the log likelihood that it belongs to each of the three seizure stages using three seperately trained HMM models. Because there are a total of $n+1$ sliding windows over $2n+1$ observations, we formed a new $3n+3$ feature vector that contains the log likelihood values for the three seizure stages over the $n+1$ sliding windows. We applied a trained classifier to this feature vector to determine the seizure stage of the spike $O_t$. For this second stage classifier, we explored the multi-class support vector machine with linear kernel and radial basis function (RBF) kernel respectively.

\subsection{Dataset and Detection Result}
We analyzed micro-electrocorticographic ($\mu$ECoG) data
from an acute in vivo feline model of epilepsy. Adult cats
were anesthetized with a continuous infusion (3 $\sim$ 10 $
mg/kg/hr$) of intravenous thiopental. Craniotomy and
durotomy were performed to expose a $2 \times 3$ $cm$ region of
cortex. The high resolution electrode array was then placed
on the surface of the brain over primary visual cortex,
localized by electrophysiological recordings of visual evoked potentials. Picrotoxin, was topically applied
adjacent to the anterior-medial corner of the electrode array
in an amount sufficient to induce abnormal electrical spikes
and seizures from the covered region\cite{delay_map1}.

\begin{table}
\centering
\caption{Description of our datasets}
\scalebox{0.8}{
    \begin{tabular}{  p{1.35cm} |  p{1.5cm} p{1.5cm} p{1.2cm}  p{1.5cm}  p{1.5cm}}
    \toprule
    Dataset & sampling frequency (Hz) & recording length (min) & labeled seizures & detected spike segments  & seizure spike segments\\ 
    \toprule
    Dataset 1 (cat 1) & 277.778  & 53.67 & 7 & 1685 & 254\\ 
    Dataset 2 (cat 2) & 925.925  & 32.33& 27 & 3706  & 3191\\ 
    Dataset 3 (cat 2) & 925.925  & 26.16 & 27 & 2380 &1930\\ 
   \bottomrule
    \end{tabular}
    }
    \begin{tablenotes}
      \small
      \item \textit{\small{Dataset 2 and Dataset 3 are from the same animal, with different implant position}}
    \end{tablenotes}
    
    \label{dataset}    
\end{table}

 \begin{table}
  \centering
      \caption{Seizure detection accuracy for three datasets using the HMM+SVM approach.}
   \scalebox{0.8}{
    \begin{tabular}{  p{2.5 cm} |  p{2 cm}  p{2.5 cm} p{2 cm} }
    \toprule
    Dataset & precision & false positive per hour & average detection delay (sec.) \\ 
    \toprule
    Dataset 1 (cat 1) & 5/7 & 0 & -1.86  \\ 
    Dataset 2 (cat 2) & 20/27 & 1.85 & -2.22   \\ 
    Dataset 3 (cat 2) & 24/27 & 0 & -2.33  \\ 
    \bottomrule
    \end{tabular}
    }
   \begin{tablenotes}
   
      \small
      \item  \textit{\small{Seizure onset is predicted if 5 consecutive spikes are classified to either pre-seizure or seizure stage. The time between this prediction and the actual onset time is defined as the delay. Seizure detection is considered correct if delay is between -8 and 0 sec. The last column reports the average detection delays. The false positive rate is higher than expected because our dataset (Table. \ref{dataset}) contains many more seizures than would be expected in naturally occurring epilepsy. }}
      %Because seizure is much more frequent than natural seizure in our dataset (Table. \ref{dataset}), the false positive per hour is higher than normal.
    \end{tablenotes}
    \label{prediction statistics datasets}    
\end{table}
The active electrode array placed on the cortex was used
to record data from 360 independent channels arranged in 20
columns and 18 rows, spaced $500 \mu m $ apart. Each electrode
contact was composed of a $300 \mu m \times 300 \mu m $ square of
platinum. Two flexible silicon transistors
for each electrode buffered and multiplexed the recorded
signals \cite{delay_map1}. The total array size was $10mm \times 9mm$. Three datasets were acquired from two different animals, as summarized in Table \ref{dataset}.

%We tested our algorithm on three recorded dataset from 2 different animal listed in the Table.\ref{dataset}. Seizure prediction with single HMM iare displayed in Table \ref {prediction statistics}.  Seizure prediction with Multiple HMMs are displayed in Table \ref{prediction statistics Multi-HMM}.
Two datasets were acquired from the same animal, with the electrode located on different areas of the cortex. The patterns observed from the second placement were distinct from the patterns observed during the first placement, motivating splitting the dataset.
We used observation sequences extracted during one seizure period, pre-seizure period (8 sec) before this seizure, and the non-seizure period leading up to this seizure as our testing data, used the rest of the data for training. We repeated this approach for each seizure.Table \ref{prediction statistics datasets} summarizes the seizure detection accuracy using the HMM+SVM approach. If the seizure actually happened after the predicted onset within 8 sec, we consider the detection accurate. Note that a large percentage of the seizure events were predicted before their onset. However, with the datasets we have, it is not possible to attempt to predict the seizures with a much longer lead-in time. This is because the seizures are induced and the intervals between successive seizures are much shorter than the interval between naturally occurring seizures. 

\section{conclusion}\label{conclusion}
$\mu$ECoG has tremendous potential for many research and clinical applications. In this work, we have combined advanced video analysis and machine learning algorithms to analyze these challenging datasets in novel ways. We have developed efficient methods for identifying and localizing the spatial and temporal extent of inter-ictal and ictal spikes through graph filtering and region growing. For those identified spikes we designed a pairwise metric that characterize the similarity between the movement patterns of different spikes. Using this similarity metric, we discovered the lower-dimension manifold features using the Isomap and furthermore discovered different spike patterns using the DPM clustering method on the manifold features. The clustering result reveals cleaner patterns that have not been seen before.

We attempted to solve a very challenging spike wavefront prediction task. We casted the wavefront prediction into a polynomial regression problem by using wavefront coordinates of the first few frame to estimate the parameters. We then used the polynomial model to predict future wavefront trajectories. 
We found that by identifying the spike cluster that a spike belongs to and requiring the polynomial coefficients to be close to the mean coefficients of the trajectories in that cluster, we are able to achieve more accurate trajectory prediction. The mean error of predicted wavefront locations is only around 4 channels away from the true locations. The error is caused mainly by two reasons. One is the error in spike segmentation using 3D region growing, which would consider merging / splitting spikes as the same one as long as these two or more spikes were spatially connected at one time. This issue makes trajectory of a spike less meaningful when there are multiple spikes that intersecting each other. The second is that polynomial fitting is one simple way of describing wave pattern, more advanced techniques like recurrent neural network or its variants could be a better model.  Another interesting issue would be how to predict spike pattern of a new spike based on its first few frames as well as the past temporal variation of spike patterns. These problems remain to be explored in our future work.

We further investigated seizure detection problem using each spike segment's cluster label. The proposed two stage HMM+SVM classifier yielded accurate seizure detection and possibly prediction, by detecting 49 out of 61 seizures before the seizure onset. Our framework has demonstrated the potential for seizure prediction by analyzing the temporal variation of the spike labels, although 
the prediction time is only within seconds of  the actual seizure onset. This is due to the fact that our datasets consisted of induced seizures that occur very frequently. As such, there does not appear to be a pre-seizure stage that occurs far ahead of the seizure onset. As a future research, we will adapt the proposed framework to datasets from more realistic seizure models or human datasets for seizure prediction.
\vspace{-0.05in}

\section{ACKNOWLEDGEMENT}
This work was funded by National Science Foundation award CCF-1422914.
 \vspace{-0.05in}

\bibliography{mybib}{}
\bibliographystyle{IEEEbib}
 
\end{document}